\documentclass[journal]{IEEEtran}
\usepackage{algorithm}
\usepackage{url}
\usepackage{amssymb}
\usepackage{amsmath}
\usepackage{epsfig}
\usepackage{epstopdf}
\usepackage{subfigure}
\usepackage{balance}
\usepackage{multirow}
\usepackage{booktabs}
\usepackage{color, colortbl}
\usepackage{tabularx}
\usepackage{algcompatible}
\usepackage{soul}

\ifCLASSINFOpdf
\else
\fi
\begin{document}
%
%\title{ThermalNet: Intelligent Building Thermal Comfort Prediction and Control with Deep Learning}
%
%\title{Learning for Intelligent Energy Optimization and Thermal Comfort Control in Smart Buildings}
%
\title{Energy-Efficient Thermal Comfort Control in Smart Buildings via Deep Reinforcement Learning}

\author{Guanyu~Gao,
        Jie~Li,
        Yonggang~Wen
\thanks{G.Y. Gao, J. Li and Y.G. Wen are with the School of Computer Science and Engineering, Nanyang Technological University, Singapore, 639798.
E-mail: \{ggao001, lijie, ygwen\}@ntu.edu.sg.}
}

\maketitle

\begin{abstract}
%
%Building electricity consumption accounts for more than 30\% of the total electricity consumption in large cities.
%
Heating, Ventilation, and Air Conditioning (HVAC) is extremely energy-consuming, accounting for 40\% of total building energy consumption.
%
%Due to the high electricity price and the environmental pollution caused by fossil fuel based electricity generation, 
%
Therefore, it is crucial to design some energy-efficient building thermal control policies which can reduce the energy consumption of HVAC while maintaining the comfort of the occupants.
However, implementing such a policy is challenging, because it involves various influencing factors in a building environment, which are usually hard to model and may be different from case to case. 
%
%including the building structures, outdoor thermal condition, and the mechanism of HVAC, etc.
%
%it is challenging to study the influence of each factor on energy consumption and thermal comfort for deriving a generally applicable thermal control policy.
%
To address this challenge, we propose a deep reinforcement learning based framework for energy optimization and thermal comfort control in smart buildings.
We formulate the building thermal control as a cost-minimization problem which jointly considers the energy consumption of HVAC and the thermal comfort of the occupants.
To solve the problem, 
we first adopt a deep neural network based approach for predicting the occupants' thermal comfort,
and then adopt Deep Deterministic Policy Gradients (DDPG) for learning the thermal control policy.
To evaluate the performance,
we implement a building thermal control simulation system and evaluate the performance under various settings.
The experiment results show that our method can improve the thermal comfort prediction accuracy, and reduce the energy consumption of HVAC while improving the occupants' thermal comfort.
\end{abstract}


\maketitle



% \fi
\IEEEpeerreviewmaketitle

\section{Introduction}\label{sec:introduction}
Building thermal control is important for providing high-quality working and living environments,
because occupants can only feel comfortable when the temperature and humidity of the indoor thermal condition are within the thermal comfort zone.
However, the ambient thermal condition may change dramatically, which will lead to the fluctuation of indoor thermal condition and the discomfort of the occupants.
Therefore, building thermal control is necessary for maintaining acceptable indoor thermal condition.
Heating, Ventilation and Air Conditioning (HVAC) systems are usually adopted for controlling indoor thermal conditions.
One main concern of the HVAC systems is their high energy consumption, which makes building energy consumption accounts for 20\%-40\% of the total energy consumption \cite{perez2008review}.
On the other hand, occupants may feel cold or hot if the set-points of the HVAC systems are inappropriate, although more energy may be consumed.
Thus, it is necessary to study  how to reduce the energy consumption of the HVAC systems while keeping occupants comfort, especially due to the high electricity price and the increase of electricity  consumption and environmental pollution.

Various factors have influences on building thermal control. These factors can be categorized into three parts, namely, HVAC system related factors, building thermal environment related factors, and human related factors.
The HVAC system can control the building thermal condition by adjusting the set-points of air temperature and humidity, which will also change the energy consumption of the HVAC system.
Building thermal environments are determined by the structures of buildings, the indoor and outdoor thermal conditions, and the heat sources (e.g., bulbs, computers, etc).
These factors affect the dynamic changes of building thermal conditions.
The feelings of the occupants under given thermal conditions are subjective and may be different from person to person \cite{corgnati2007perception},
however, the degrees of occupants' satisfaction under given thermal conditions need to be precisely estimated to select the appropriate set-points of the HVAC system for efficient thermal control.
The factors of these three parts are correlated, and we need to consider these factors altogether for achieving thermal comfort and reducing energy consumption.

Many approaches have been proposed for building thermal control and energy optimization \cite{dounis2009advanced}.
The model-based approaches aimed to model the thermal dynamics with simplified mathematical models, such as Proportional Integrate Derivative (PID)  \cite{levermore1992building, dounis1996comparison},
Model Predictive Control (MPC) \cite{shepherd2003fuzzy, calvino2004control},
Fuzzy Control \cite{shepherd2003fuzzy, calvino2004control},
Linear-Quadratic Regulator \cite{maasoumy2011model}.
However, the complexity of the thermal dynamics and the various influencing factors are hard to be precisely modelled by these methods.
These methods also require customized design for the control policy for specified environments.
Another line of works adopted the learning-based approach for learning the optimal control policy,
such as the reinforcement learning approach.
The reinforcement learning approach can lean the optimal control policy by interactions with the thermal environment.
However, the plain reinforcement learning approaches (e.g., \cite{barrett2015autonomous}, \cite{li2015multi}, \cite{nikovski2013method, zenger2013towards, fazenda2014using, dalamagkidis2007reinforcement})
cannot achieve high performance if the state-action space is large.
Deep reinforcement learning approaches (e.g., \cite{wei2017deep, wang2017long}) have been recently adopted to address this limitation by using deep learning for representation of the large state-action space.
However, the methods adopted in \cite{wei2017deep, wang2017long} require discretization of the state-action space, which will also limit the performance of the thermal control policy.

Different from previous works, we propose a learning-based framework for joint energy optimization and thermal comfort control in smart buildings. 
We first design a deep neural network method with Bayesian regularization for predicting the occupants' thermal comfort by considering different influencing factors.
The method can predict the occupants' thermal comfort precisely, therefore, the thermal comfort prediction results can be used as a feedback for thermal control.
Then, we adopt deep reinforcement learning approach for thermal control to minimize the overall cost by jointly considering energy consumption of the HVAC system and the thermal comfort of the occupants.
We adopt Deep Deterministic Policy Gradients (DDPG) approach \cite{lillicrap2015continuous}, which is a reinforcement learning algorithm for continuous control problem, for learning the optimal control policy.
This is because in thermal control, temperature, humidity, and some other control variables are all continuous, therefore, DDPG is very suitable for addressing the problem in this scenario.
Compared with the methods adopted in \cite{wei2017deep, wang2017long}, DDPG can avoid the discretization of the control variables (e.g., temperature, humidity), which can also improve the control precision.
To evaluate the performance of our method, we implement a thermal control simulation system with TRNSYS and evaluate the performance under various settings.
The main contributions of this paper are as follows:
\begin{itemize}
  \item We propose a DDPG approach for learning the control policy for energy optimization and thermal comfort control. Our method can improve the occupants' thermal comfort and reduce the energy consumption of HVAC.

  \item We design a deep neural network method with Bayesian regularization for predicting thermal comfort. Our method can predict thermal comfort more accurately.

  \item We implement a building thermal control simulation system with TRNSYS, and conduct extensive experiments for evaluating the performances of our proposed method under different experiment settings.
\end{itemize}

The rest of this paper are organized as follows.
Section \ref{sec:Background-Related} presents the background and related works.
Section \ref{sec:system-design} introduces the system design of the thermal control system in smart buildings.
Section \ref{sec:system-model} presents the system model and problem formulation.
Section \ref{sec:learning-algo} presents the deep neural network based method for predicting thermal comfort and the DDPG method for learning the control policy.
Section \ref{sec:performace-evaluation} evaluates the performance of our method and compare it with the baseline methods.
Section  \ref{sec:conclusion} concludes this paper.

\section{Background and Related Work} \label{sec:Background-Related}
In this section, we first introduce the background and then discuss the related works on building thermal control.

\subsection{Background}
\subsubsection{HVAC}
The HVAC system \cite{osti_6352074} is used to provide thermal comfort and maintain indoor air quality in buildings.
The main functionalities of the HVAC system are heating, ventilation, and air conditioning.
Specifically, heating is to generate heat to raise the air temperature in the building.
Ventilation is to exchange air with the outside and circulate the air within the building.
Air conditioning provides cooling and humidity control.
In this work, we mainly study the set-points and the energy consumption of the HVAC system, 
and we do not consider the inside mechanisms of the HVAC system (i.e., how the set-points are achieved by the HVAC system via its inside functioning),
because different HVAC system may have different inside mechanisms.
We design the algorithm independent of the inside mechanisms so that it can be generally applied in different types of HVAC systems.

\subsubsection{Thermal Comfort}
Thermal comfort reflects the occupants' satisfaction with the thermal condition \cite{american1992ashrae}.
For quantitatively evaluating thermal comfort, thermal comfort models are introduced for predicting the occupants' satisfaction under given thermal conditions.
Because the occupants' feelings under given thermal conditions are subjective,
thermal comfort is assessed by subjective evaluation.
Many subjects will be invited to evaluate their degrees of satisfaction under different thermal conditions, such as cold (-2), cool (-1), neutral (0), warm (1), and hot (2).
Then, some mathematical or heuristic methods can be adopted for fitting the data.
Many models have been developed for evaluating the thermal comfort of the occupants under different thermal conditions \cite{cheng2012thermal}, for instance, Predicted Mean Vote (PMV), Actual Mean Vote (AMV), Predicted Percentage Dissatisfied (PPD), etc.
%
% Among these models, Predicted Mean Vote (PMV) model \cite{fanger1970thermal} is currently the most recognized thermal comfort model.
%

\subsubsection{Reinforcement Learning}
Reinforcement learning is concerned with how the agent should take actions in a dynamic environment to maximize its overall rewards \cite{van2012reinforcement}.
It can learn the optimal control policy by trails during the interactions with the environment.
The reinforcement learning process can be modelled as a Markov Decision Process (MDP), which consists of State, Action, and Rewards, etc.
The agent first observes the current state of the environment and select an action to take according to a certain policy, followed by obtaining the rewards for the action taken on the observed state.
These steps will be iterated during the learning stage and the control policy will be updated until it is converged.
The optimal policy can be learned during the trails without directly modeling the system dynamics, therefore, reinforcement learning is a natural solution if the policy can be learnt by the interactions with the environment  and the dynamics of the system is hard to model precisely.
Deep reinforcement learning \cite{mnih2015human} is a combination of deep learning and reinforcement learning.
It can significantly improve the performance compared with reinforcement learning if the dimensions of state and action is large, because deep learning has higher capacity for learning the representation of large space and tremendous data.
\subsection{Related Work}
The existing mathematical approaches for building thermal control can be generally classified into two categories, namely, model-based approaches and learning-based approaches.
Specifically, the model-based approaches derive the policy by modeling the dynamics of the environment;
the learning-based approaches derive the policy by learning from the interactions with the environment.

\subsubsection{Model-based Approaches}
Levermore \emph{et al.} \cite{levermore1992building} and Dounis \emph{et al.} \cite{dounis1996comparison} used the Proportional Integrate Derivative (PID) method for building energy management and indoor air quality control.
Shepherd \emph{et al.} \cite{shepherd2003fuzzy} and Calvino \cite{calvino2004control} proposed a fuzzy control method for managing building thermal conditions and energy cost.
Kummert \emph{et al.} \cite{kummert2001optimal} and Wang \emph{et al.} \cite{wang2000model} proposed the optimal control method for controlling the HVAC system.
Ma \emph{et al.} \cite{ma2012model} introduced a Model Predictive Control (MPC) based approach for controlling building cooling systems by considering thermal energy storage.
Wei \emph{et al.} \cite{wei2014co} adopted the MPC based approach for jointly scheduling HVAC, electric vehicle and battery usage for reducing building energy consumption while keeping temperature within comfort zone.
Maasoumy \emph{et al.} \cite{maasoumy2011model} proposed a tracking linear-quadratic regulator for balancing human comfort and energy consumption in buildings.
Oldewurtel \emph{et al.} \cite{oldewurtel2010energy} proposed a bilinear model under stochastic uncertainty for building climate control while considering weather predictions.
These model-based approaches strive to model the dynamics of the thermal environment with some simplified mathematical models.
However, the thermal environment is affected by various factors and is complicated in nature, and it is hard to be precisely modeled.
Moreover, the performances of the model-based approaches are constrained by the specified building environments, and it is hard to derive a generalized model-based approach that can be applicable in various building environments.

\subsubsection{Learning-based Approaches}
Barrett \emph{et al.} \cite{barrett2015autonomous}, Li \emph{et al.} \cite{li2015multi} and Nikovski \emph{et al.} \cite{nikovski2013method} adopted Q learning based approaches for the HVAC control.
Zenger \emph{et al.} \cite{zenger2013towards} adopted State-Action-Reward-State-Action (SARSA) for achieving the desired temperature while reducing energy consumption.
Fazenda \emph{et al.} \cite{fazenda2014using} proposed a neural fitted reinforcement learning approach for learning how to schedule thermostat temperature set-points.
Dalamagkidis \emph{et al.} \cite{dalamagkidis2007reinforcement} designed the Linear Reinforcement Learning Controller (LRLC) using linear function approximation of the state-action value function to achieve thermal comfort with minimal energy consumption.
Anderson \cite{anderson2004robust} proposed a robust control  framework for combined Proportional Integral (PI) control and reinforcement learning control for HVAC of buildings.
Wei \emph{et al.} \cite{wei2017deep} adopted a neural network based deep Q learning method for the HVAC control.
Wang \emph{et al.} \cite{wang2017long} adopted a Long-Short Term Memory (LSTM) recurrent neural network based reinforcement learning controller for controlling air conditioning system.
The tabular Q learning approach, SARSA, and other plain reinforcement learning approaches adopted in \cite{barrett2015autonomous, li2015multi, nikovski2013method, zenger2013towards, fazenda2014using, dalamagkidis2007reinforcement, anderson2004robust}  are not suitable for problems with large state-action spaces,
partly due to that plain reinforcement learning approaches fail to achieve satisfying generalization of the value function and policy function in large spaces.
Deep Q learning \cite{wei2017deep} and LSTM based reinforcement learning \cite{wang2017long} can improve the performances with neural networks, which have better generalization capacity.
However, the proposed approaches in these works require discretization of the state-action space, which will decrease the control precision and the performance.
To address these drawbacks in previous works, we propose the DDPG approach for continuous control with deep reinforcement learning in the HVAC system.
Our method also provides the capacity for customizing the thermal comfort settings by jointly considering energy consumption and thermal comfort. 
One can customize the thermal comfort threshold according to the occupants' thermal requirements for reducing the energy consumption of the HVAC system.

\section{System Design and Control Flow} \label{sec:system-design}
In this section, we present the system design and the workflow for thermal control in smart buildings.

\begin{figure}
\begin{center}
\epsfig{file=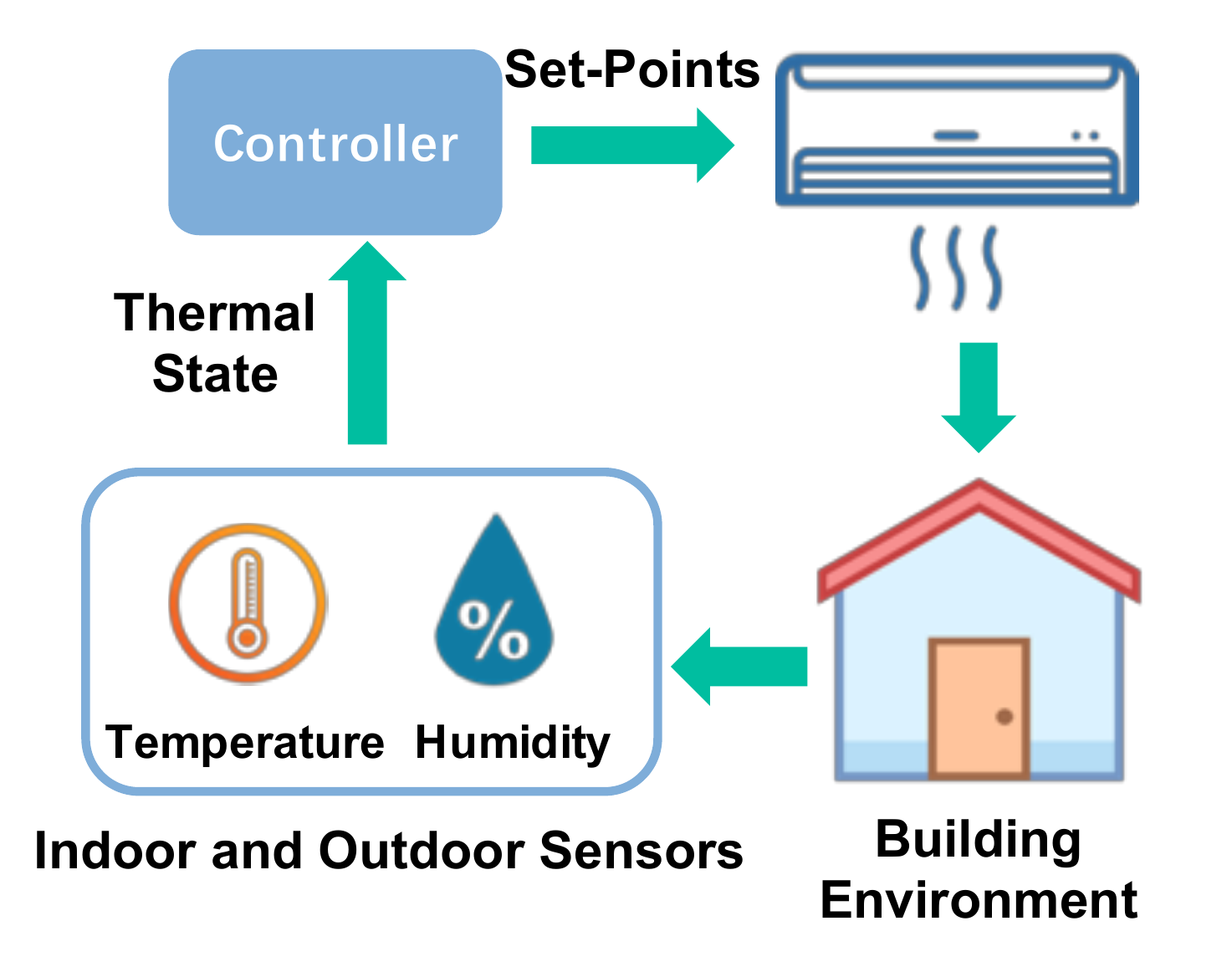, width=0.75\columnwidth}
\end{center}
%\vspace{-3 mm}
\caption{The system design of the thermal control system. The controller obtains the building thermal state information from the sensors and make control actions by adjusting the set-points of the HVAC system.} \label{fig-system-design}
\end{figure}

\subsection{System Design}
The design of the thermal control system is illustrated in Fig. \ref{fig-system-design}.
The system consists of the following components, namely, the sensors, the controller, and the HVAC system.
The functionalities of each component are detailed as follows.

\emph{Sensors:}
The sensors periodically measure the thermal conditions of the indoor and outdoor building environments \cite{pan2015internet},
including temperature, humidity, etc.
The sensors are connected with the controller via Internet of things (IoT) networks \cite{minoli2017iot},
and the sensors will send the collected information to the controller for making thermal control decisions.

\emph{Controller:}
The controller collects the thermal state information of the indoor and outdoor building environment and the energy consumption and the working state information of the HVAC system \cite{pan2015internet}.
Based on these information, the controller will take control actions by updating the set-points of the HVAC system periodically according to the control policy.

\emph{HVAC:}
The HVAC system will function according to the set-points of the controller.
For instance, if the set-point of the temperature {}is lower than the current indoor temperature,
the HVAC system will start cooling until the indoor temperature matches with the set-point temperature.
If the set-point temperature is higher than the current indoor temperature,
the HVAC system will start heating until achieving the specified indoor temperature.
The HVAC system may also need to constantly keep heating, cooling, or ventilation for keeping the specified indoor thermal condition due to the influence of the outside building thermal environment.
\subsection{Control Flow}
\begin{figure}
\begin{center}
\epsfig{file=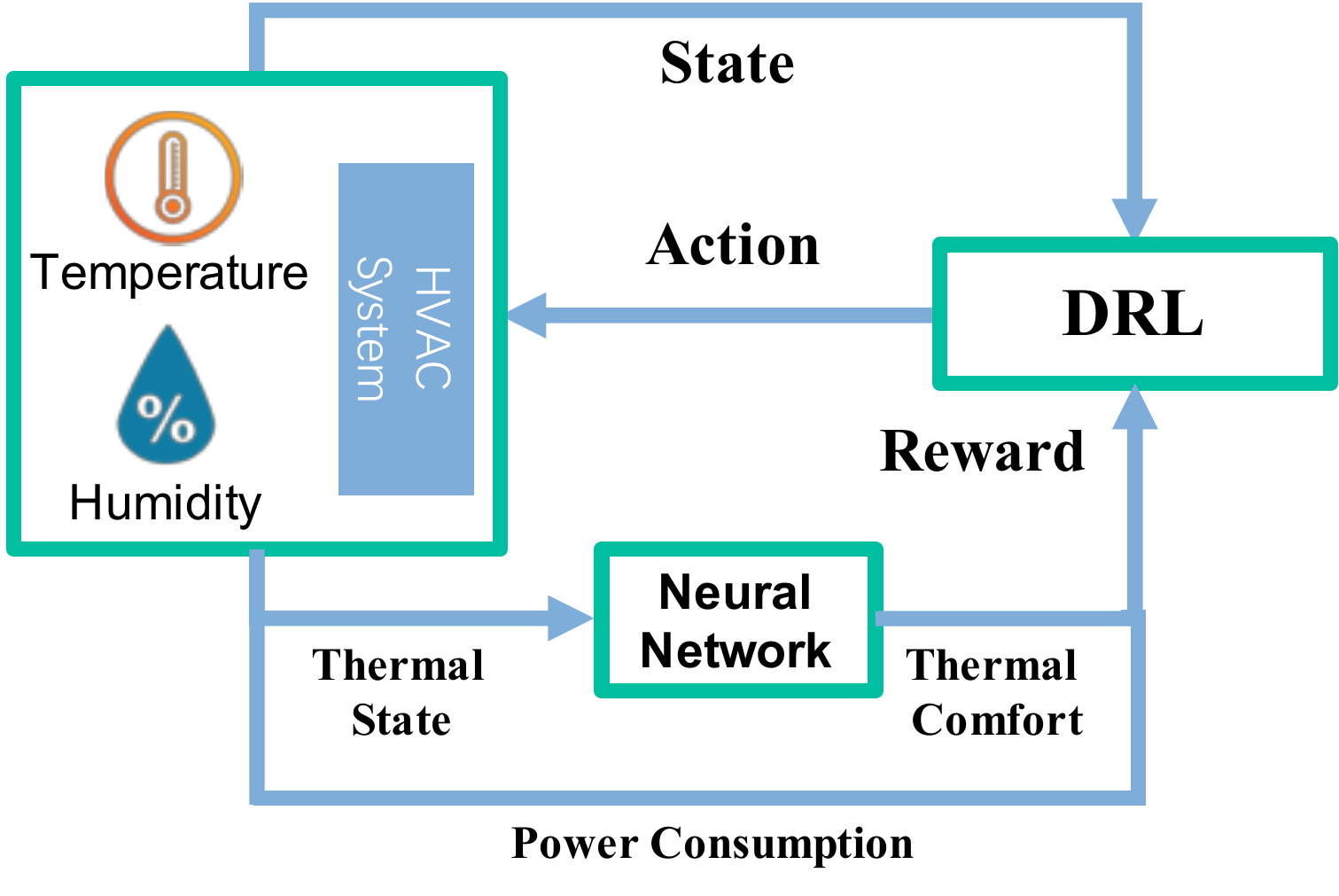, width=0.75\columnwidth}
\end{center}
%\vspace{-3 mm}
\caption{The control flow of the building thermal control system. We adopt deep neural network for thermal comfort prediction and Deep Reinforcement Learning (DRL) for thermal control.} \label{fig-control-logic}
\end{figure}
We illustrate the control flow of the system in Fig. \ref{fig-control-logic}.
We adopt deep neural network  for predicting the occupants' thermal comfort value based on the current indoor thermal state, and we use deep reinforcement learning for making control decisions for thermal control.
The neural network based thermal comfort predictor can be trained offline using existing thermal prediction dataset.
After training, the indoor building thermal state information will be input into the trained thermal prediction model for predicting thermal comfort value.
The thermal comfort prediction value and the energy consumption information of the HVAC system will be used for calculating the reward during each time slot.
The deep reinforcement learning based thermal controller can learn the control policy by observing the received rewards for taking actions on different states.
We train the thermal controller using the building thermal control simulation system, which will be detailed in Section \ref{sec:thermal-control-implementation}.
After training, the indoor and outdoor building thermal state will be input into the trained thermal control model for making thermal control actions.

\section{Problem Formulation} \label{sec:system-model}
In this section, we introduce the notations and formulate the energy optimization and thermal comfort control as a cost-minimization problem using MDP.
We adopt a discrete-time model and the time slot is denoted as $t=0,1,2,...$.
The duration of each time slot is from several minutes to one hour. 
The notations used in this paper are summarized in Table \ref{tabel_notation}.

\subsection{Building Thermal State}
The energy consumption of the HVAC system is affected by both of the indoor thermal environment and the outdoor thermal environment.
For the indoor and outdoor thermal environments, we consider the factors of  air temperature and humidity, which have the greatest influences on the energy consumption of the HVAC system and the human thermal comfort.
We denote the indoor air temperature and humidity at time slot $t$ as $T^{in}_{t}$ and $H^{in}_{t}$, respectively;
and we denoted the outdoor air temperature and humidity at time slot $t$ as $T^{out}_{t}$ and $H^{out}_{t}$, respectively.
The indoor and outdoor air temperature and humidity can be obtained from the sensors in a smart building.

\subsection{Set-Points of HVAC System}
The controller can change the set-point air temperature and humidity of the HVAC system for controlling the indoor thermal condition.
We denote the set-point air temperature of the HVAC system at time slot $t$ as $T^{set}_{t}$, and denote the set-point air humidity as time slot $t$ as $H^{set}_{t}$.
At the beginning of each time slot, the controller updates the set-point air temperature and humidity of the HVAC system according to the indoor and outdoor thermal condition for controlling thermal comfort and energy consumption.
%
%We consider the HVAC system as a black box system. The inputs are the set-point air temperature and humidity of the HVAC system,
%
%and the outputs are heating, ventilating, or cooling.
%

\begin{table}
\centering
\caption{Key Notation and Definition} \label{tabel_notation}
\begin{tabular}{p{0.8cm} p{7.0cm}} \hline \hline
$t$                 & the discrete time slot, $t=0,1,2,...$ \\
$T^{in}_{t}$        & the indoor air temperature at time slot $t$ \\
$H^{in}_{t}$        & the indoor air humidity at time slot $t$ \\
$T^{out}_{t}$       & the outdoor air temperature at time slot $t$ \\
$H^{out}_{t}$       & the outdoor air humidity at time slot $t$ \\
$T^{set}_{t}$       & the set-point air temperature at time slot $t$\\
$H^{set}_{t}$       & the set-point air humidity at time slot $t$\\
$M_t$               & predicted thermal comfort value at time slot $t$\\
$\Phi$              & thermal comfort prediction algorithm \\
$P_t$               & the energy consumption of the HVAC system at time slot $t$\\
$S_t$               & the state of the MDP at time slot $t$\\
$A_t$               & the action of the MDP at time slot $t$\\
$R_t$               & the reward of the MDP at time slot $t$\\
$\pi$               & the thermal control policy\\
$\gamma$            & the discount factor for the reward\\
$D$                 & the threshold for thermal comfort\\
$\beta$              & the weight of the penalty for energy consumption\\
$\alpha_1, \alpha_2$ & Bayesian hyperparameters \\
$n$                  & the number of training samples \\
$m$                  & the number of weights in the neural network\\
$w_j$                & the $j$-th weight in the neural network \\
$\theta^{Q}, \theta^{\mu}$ & the parameters of the critic network and the actor network \\
$N(t)$               & the exploration noise for training\\ 
$\tau$               & the discount factor for model update \\
\hline \hline\end{tabular}
\end{table}

\subsection{Thermal Comfort Prediction}
Thermal comfort prediction predicts the occupants' satisfaction with the thermal condition.
Many factors may influence the occupants' thermal comfort, e.g., metabolic rate, clothing insulation, air temperature and humidity, skin wetness, etc.
For some of the factors, they may be different from person to person, or cannot be easily measured in real environments,
such as the occupants' metabolic rate and skin wetness.
In this work, we mainly consider the indoor air temperature and humidity as variables for thermal comfort prediction,
and the other factors are set as default values.
The occupants' thermal comfort value at time slot $t$ is predicted as
\begin{equation}\label{equ-thermal-comfort}
M_t = \Phi(T^{in}_{t}, H^{in}_{t}),
\end{equation}
where $M_t$ is the predicted thermal comfort value and $\Phi$ is the thermal comfort prediction algorithm.
In this work, we adopt the deep neural network method for predicting thermal comfort value,
which will be detailed in Section \ref{sec:thermal-comfort-prediction}.

\subsection{Energy Consumption of HVAC system}
The HVAC system will consume energy for heating, cooling, and dehumidification.
The information of the energy consumption of the HVAC system during each time slot can be obtained from the smart meter.
We assume that the energy consumption information of the HVAC system during each time slot is known, and we denote the energy consumption of the HVAC system at time slot $t$ as $P_t$.
To make the modeling independent of the inside mechanisms of the HVAC system, we only consider the overall energy consumption of the HVAC system during each time slot. The specified energy consumptions for heating, cooling, dehumidification, or other inside mechanisms are considered as unknown.

\subsection{Problem Formulation}
We formulate the energy optimization and thermal comfort control in the smart building as a MDP process for minimizing the overall cost over time, the details of which are as follows.

\textbf{State:}
The state of the MDP is the current indoor and outdoor thermal conditions at each time slot, represented as
\begin{equation}\label{equ-mdp-state}
S_t = (T^{in}_{t}, H^{in}_{t}, T^{out}_{t}, H^{out}_{t}),
\end{equation}
where $S_t$ is the state of the MDP at time slot $t$.

\textbf{Action:}
The action of the MDP is the set-point air temperature and humidity of the HVAC system, represented as
\begin{equation}\label{equ-mdp-action}
A_t = (T^{set}_{t}, H^{set}_{t}),
\end{equation}
where $A_t$ is the action of the MDP at time slot $t$.
The action is determined according to the control policy by observing the current state. Mathematically, it can be represented as
\begin{equation}
A_t = \pi(S_t),
\end{equation}
where $\pi$ is the control policy for thermal control.

\textbf{Reward:}
The reward of the MDP consists of two parts, namely, the penalty for the energy consumption of the HVAC system and the penalty for the occupants' thermal discomfort.
Specifically, the reward should be less, if more energy is consumed by the HVAC system or the occupants feel uncomfortable about the building thermal condition.
The predicted value of thermal comfort ($M_t$) ranges from -3 to 3, where -3 is too cold and +3 is too hot and 0 is neutral.

The occupants can feel comfort when the predicted thermal comfort value is within an acceptable range.
We denote the range as $[-D, D]$, where $D$ is the threshold for thermal comfort value.
If the predicted thermal comfort value lies within $[-D, D]$, it will not incur penalty because the occupants feel comfortable.
Otherwise, it will incur penalty for the occupants' dissatisfaction with the building thermal condition.
By jointly considering the predicted thermal comfort value and the energy consumption of the HVAC system, we calculate the overall reward of the MDP during a time slot as
\begin{equation}\label{eqn:reward-function}
R_t(S_t, A_t) = - \beta P_t -
\begin{cases}
0                & -D < M_t < D\\
(M_t - D)   & M_t > D  \\
(-D - M_t)   & M_t < -D \\
\end{cases},
\end{equation}
where $R_t$ is the overall reward for time slot $t$ and $\beta$ is the weight of the energy consumption of the HVAC system.
The weigh, $\beta$, reflects the relative importance of the energy consumption of the HVAC system compared to the occupants' thermal comfort.
If the occupants' thermal comfort is more important, $\beta$ should be set as a smaller value.
Otherwise, $\beta$ should be set as a larger value for achieving energy-efficiency.

\textbf{Cost Minimization:}
Our objective is to maximize the overall discount rewards from the current time slot by deriving the optimal thermal control policy for energy optimization and thermal comfort control,
mathematically, the objective is 
\begin{equation}\label{equ-optimization-objective}
\max_{\pi} \sum_{t'=0}^{\infty} \gamma^{t'} R_{t+t'}(S_{t+t'}, A_{t+t'}),
\end{equation}
where $\gamma$ is the discount factor.
If the precise transitions of the system states are available, we can derive the optimal policy offline.
However, it is impossible to obtain the state transition probabilities in such a complex system, therefore, the optimal control policy cannot be derived directly.
It motivates us to adopt the learning algorithm to learn the optimal policy.

\begin{figure}
\begin{center}
\epsfig{file=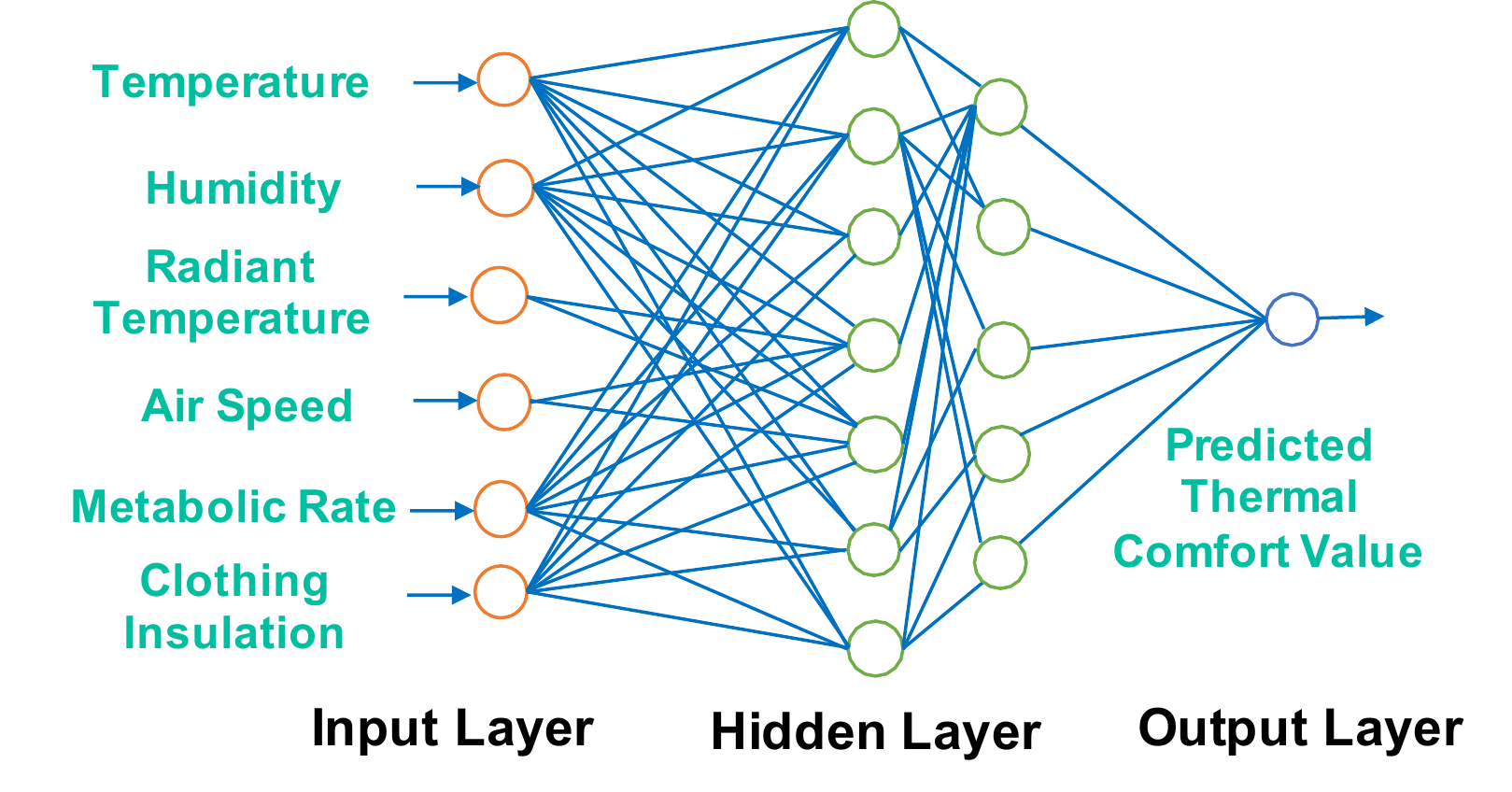, width=0.8\columnwidth}
\end{center}
\vspace{-3 mm}
\caption{The structure of the deep neural network for predicting thermal comfort. The inputs of the neural network include temperature, humidity, radiant temperature, air speed, metabolic Rate, and clothing insulation. The output of the neural network is the predicted thermal comfort value.
 } \label{fig-neural-network}
\end{figure}

\section{Learning Algorithm for Thermal Comfort Prediction and Thermal Control} \label{sec:learning-algo}
In this section, we first introduce the deep neural network based method for thermal comfort prediction.
Then, we introduce the DDPG method for learning thermal control policy.

%\subsection{Thermal Comfort Prediction} \label{sec:learning-predict-algo}
%
\subsection{Deep Neural Network based Thermal Comfort Prediction} \label{sec:thermal-comfort-prediction}
We adopt the feed-forward neural network for predicting thermal comfort.
The structure of the neural network that we adopt for predicting thermal comfort is illustrated in Fig. \ref{fig-neural-network}.
%
%It consists of three types of layers, namely, input layer, hidden layer, and output layer.
%
The inputs of the neural network include air temperature, humidity, mean radiant temperature, air speed, metabolic rate, clothing insulation, and all of these values are numerical.
The hidden layer of the neural network has two layers, and the output layer has one neuron.
The output of the neural network is the predicted thermal comfort value.
The activation function of the hidden layer is sigmoid function, and the activation function of the output layer is a linear function.

For training the neural network, some thermal  comfort prediction datasets should be adopted as the training data.
These datasets are labeled by the subjects for evaluating their thermal comfort under different thermal conditions,
and the labeled data can be noisy.
To interpolate the noisy data, we adopt Bayesian regularization \cite{foresee1997gauss} for avoiding overfitting.
The cost function for training the neural network with Bayesian regularization is to minimize the training error using the minimal weights of the neural network, represented as 
\begin{equation}\label{eqn:Bayesian-cost-function}
Cost \ Function = \alpha_1 \sum_{i=1}^{n}(Y_i - Y_i^{'})^2 + \alpha_2 \sum_{j=1}^{m} w_j^2,
\end{equation}
where $\alpha_1$ and $\alpha_2$ are Bayesian hyperparameters for specifying the  direction of the learning process to seek (i.e., minimize error or  weights),
$n$ is the number of training samples, $Y_i$ is the $i$-th labeled value by the subject, $Y_i^{'}$ is the predicted value by the neural network,
$m$ is the number of weights in the neural network, and $w_j$ is the $j$-th weight.
The training algorithm is Levenberg-Marquardt backpropagation \cite{hagan1994training}.

\subsection{Learning Thermal Control Policy with DDPG}
%
%In this section, we present the algorithm for learning the thermal comfort control policy with DDPG.
%
%We first present why we choose DDPG and then present the online algorithm design with DDPG.

\subsubsection{Why DDPG?}
\begin{figure}
\begin{center}
\epsfig{file=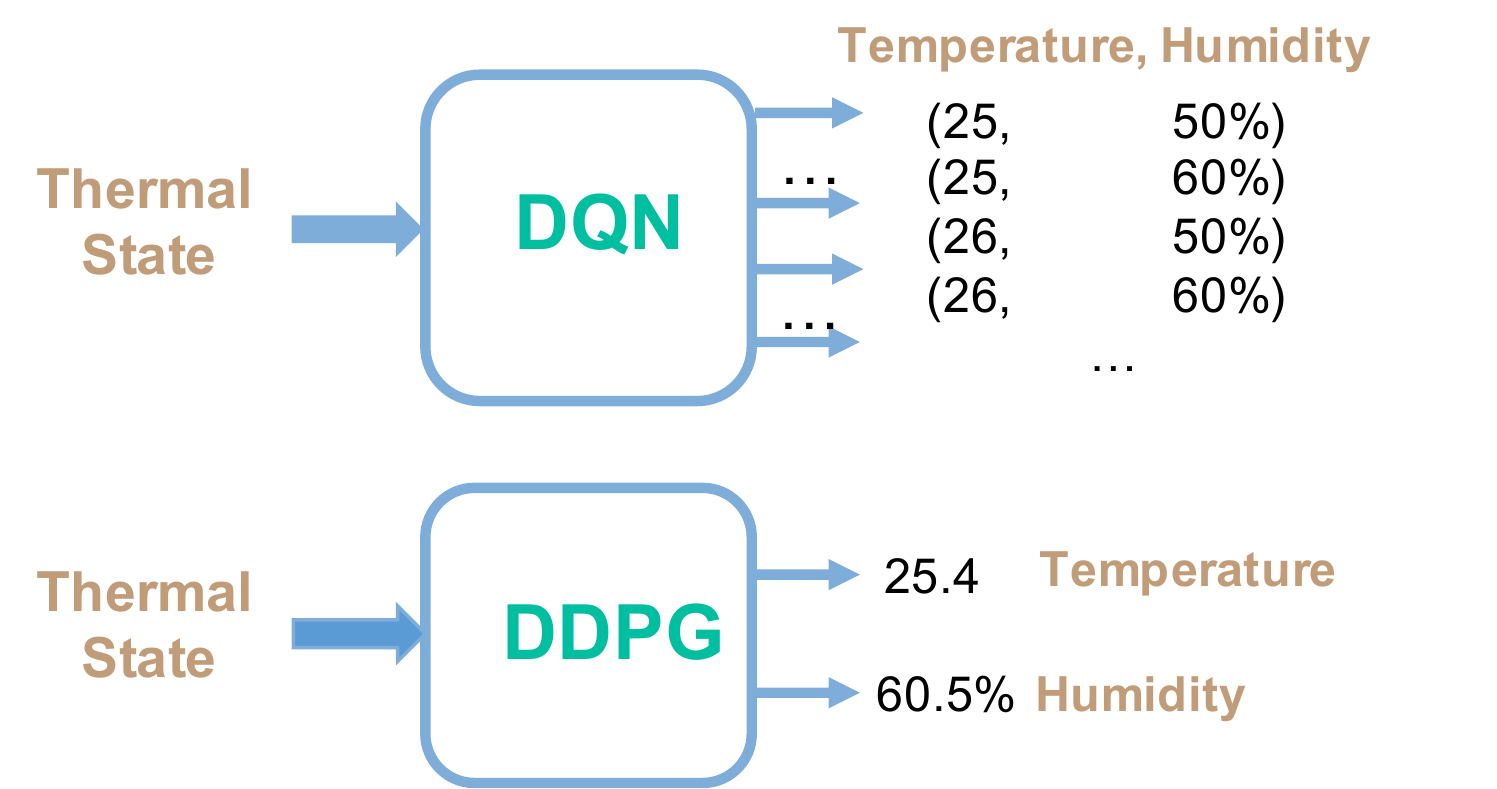, width=0.75\columnwidth}
\end{center}
\vspace{-3 mm}
\caption{A comparison of DQN and DDPG for thermal control. Temperature and humidity are continuous values, DDPG can be directly applied for continuous action control and DQN needs discretization of the action space.} \label{fig-DQN-DDPG}
\end{figure}
The control variables of the HVAC system (i.e., air temperature and humidity) take continuous values.
In Q learning and Deep Q learning, it can only handle problems with discrete and low-dimensional action spaces \cite{lillicrap2015continuous}.
Therefore, DQN cannot be directly applied to the thermal control problem, because it needs to find the control action, which is continuous, for maximizing the action-value function.
For applying DQN in thermal control, one obvious approach is to discretize the action space.
However, if one wants to achieve finer grained dicretization, it may lead to an explosion of the number of actions.
For instance, the range of humidity is from 0 to 100, if we discretize with the granularity of one, it will turn into 101 actions.
Similarly, suppose that the range of the air temperature is from 15 to 35, and there will be 200 actions if the granularity is 0.1.
If one output of DQN represents one possible combination of temperature value and humidity value, the DQN network will have 20,200 outputs.
Therefore, it will lead to a large number of outputs of DQN , which may require more training data and decrease the performance.

Compared with Q-Learning and DQN, DDPG is a natural solution for thermal control, because the action space of DDPG is continuous, and we can directly obtain the set-points of the HVAC system from the outputs of DDPG.
We illustrate the comparisons of DQN and DDPG in Fig. \ref{fig-DQN-DDPG}.
With DDPG, the network only has two outputs, namely, the real-valued set-points of air temperature and humidity of the HVAC system.
In contrast, DQN may require thousands of outputs, and each output is a possible combination of the set-points of the temperature and humidity of the HVAC system.
\subsubsection{Learning Control Policy}
\begin{figure}
\begin{center}
\epsfig{file=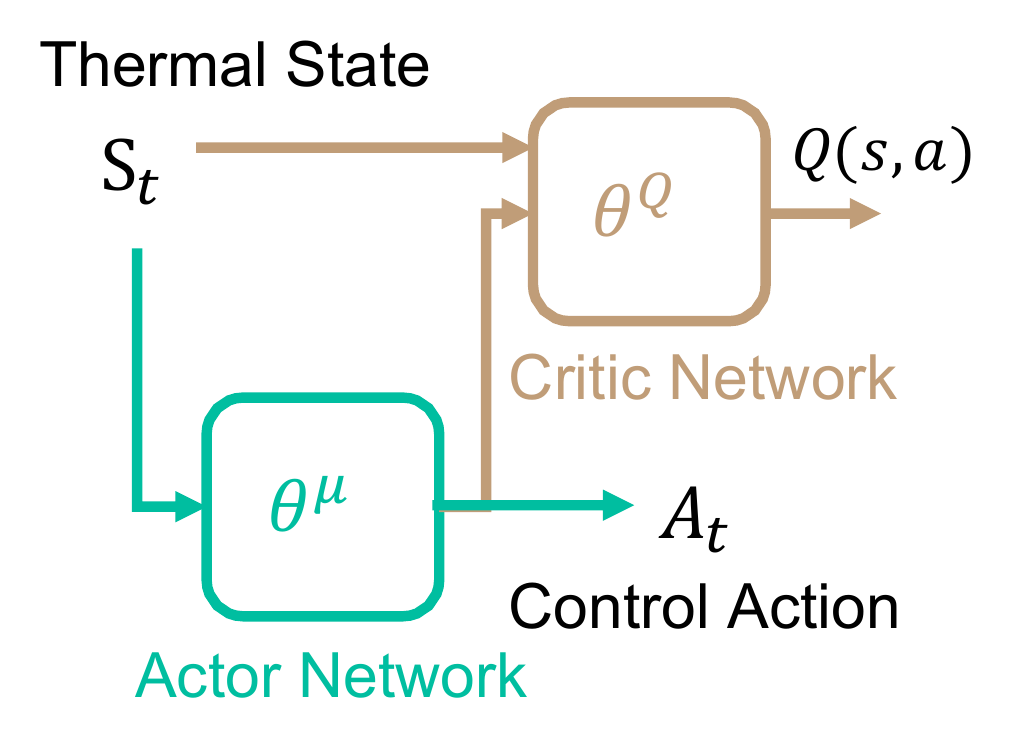, width=0.55\columnwidth}
\end{center}
\vspace{-3 mm}
\caption{The network architecture of DDPG. The actor network specifies a control action given the current thermal state and the critic network outputs an evaluation of the action generated by the actor network.} \label{fig:ddpg-framework}
\end{figure}

%
%\subsubsubsection{Network Architecture of DDPG}
%
DDPG adopts an actor-critic framework based on Deterministic Policy Gradient (DPG) \cite{silver2014deterministic}.
We illustrate the network architecture of DDPG in Fig. \ref{fig:ddpg-framework}.
The actor network is denoted as $A_t = \mu(S_t|\theta^\mu)$, where $S_t$ is the thermal state and $\theta^\mu$ represents the weights of the actor network and $A_t$ represents the control action.
The actor network maps the thermal state to a specific control action (i.e., the set-points of the HVAC system).
The critic network is denoted as $Q(S_t, A_t|\theta^Q)$, where $A_t$ is the specified control action by the actor network and $\theta^Q$ represents the weights of the critic network.
The action-value function, $Q(S_t, A_t|\theta^Q)$, describes the expected reward by taking action $A_t$ at state $S_t$ by following the policy.
The actor network and the critic network will be trained based on Temporal Difference (TD) error.
The control action specified by the actor network will be used for selecting actions during the training.
After training, only the actor network is required for making control actions.

\begin{algorithm}
\renewcommand{\algorithmicrequire}{\textbf{Input:}}
\renewcommand\algorithmicensure {\textbf{Output:} }
\caption{Training Thermal Control Policy with DDPG} \label{algo:ddpg-learning-algo}
\begin{algorithmic}[1]
\STATE{Initialize critic network $Q(S_t, A_t|\theta^Q)$ and actor network $\mu(S_t|\theta^\mu)$
with random weights $\theta^Q$ and $\theta^\mu$}
\STATE{Initialize target network $Q'(S_t, A_t|\theta^{Q'})$ and actor network $\mu'(S_t|\theta^{\mu'})$ with $\theta^{Q'} \leftarrow \theta^Q$ and $\theta^{\mu'} \leftarrow \theta^\mu$}
\STATE{Initialize replay buffer B}
\FOR{episode = 0,1,...,M}
\STATE{Obtain the initial thermal state $S_0$}
\FOR{t = 0,1,...,T}
\STATE{Obtain control action $A_t$ according to Eq. \eqref{eqn:action-with-noise}
\STATE{Update the set-points of the HVAC system according to the control action, $A_t$}
\State{Obtain new thermal state $S_{t+1}$ and calculate reward $R_t$ according to Eq. \eqref{eqn:reward-function} at the end of time slot $t$}
\State{Store $(S_t, A_t, R_t, S_{t+1})$ into replay buffer $B$}
\State{Randomly sample N transitions from replay buffer $B$}
\State{Calculate the estimated reward for each sampled transition using Eq. \eqref{eqn:estimated-reward}}
\State{Update the critic network by minimizing the MSE over the sampled minibatch and update the actor network using the sampled policy gradient}
\State{Update target network $Q'$ and $\mu'$ using Eq. \eqref{eqn:update-target-network}}
\ENDFOR
\ENDFOR
}%
\end{algorithmic}
\end{algorithm}

The training for the DDPG network is performed by interacting with the building thermal environment.
The training procedure is illustrated in Algorithm \ref{algo:ddpg-learning-algo}.
At the beginning of each time slot $t$, we first obtain the current indoor and outdoor thermal state $S_t$,
and input the thermal state into the policy network, which will output the control action.
During the training, we need to explore the state space so that the policy will not converge to local optimal solutions.
Therefore, we add a random noise to the obtained control action for exploration,
\begin{equation}\label{eqn:action-with-noise}
A_t = \mu(S_t|\theta^\mu) + N(t),
\end{equation}
where $N(t)$ is the exploration noise and $A_t$ is the control action added with exploration noise.
In our work, we use an Ornstein-Uhlenbeck process \cite{uhlenbeck1930theory} for generating the noise, $N(t)$, for exploration.
Then, the control action $A_t$ will be applied to the HVAC system.
At the end of time slot $t$, we will obtain the new thermal state $S_{t+1}$ and calculate the overall reward $R_t$ during the time slot.
The transition $(S_t, A_t, R_t, S_{t+1})$ will be stored in the replay buffer $B$ for training the policy network and the actor network.
After that, we will randomly sample $N$ transitions from the replay buffer for training the network.
For each transition $(S_i, A_i, R_i, S_{i+1}) \in N$, 
the estimated reward is calculated as follow,
\begin{equation}\label{eqn:estimated-reward}
R_i^{'} = R_i + \gamma Q'(S_{i+1}, \mu'(S_{i+1}|\theta^{\mu'})|\theta^{Q'}).
\end{equation}
The critic network will be updated by minimizing the MSE between the estimated reward ($R_i^{'}$ calculated from Eq. \eqref{eqn:estimated-reward}) and the reward predicted by the critic network ($Q(S_i, A_i)$) 
over the sampled minibatch, and the actor network will be updated using the sampled policy gradient \cite{lillicrap2015continuous}.
Then, the target networks will be updated using the following equations,
\begin{equation}\label{eqn:update-target-network}
\theta^{Q'} \leftarrow \tau \theta^{Q} + (1-\tau) \theta^{Q'}, \theta^{\mu'} \leftarrow \tau \theta^{\mu} + (1-\tau) \theta^{\mu'},
\end{equation}
where $\tau$ is the discount factor for model update.
After training, only the actor network is applied for making control action.
The set-points of the HVAC system will be set as the control actions specified by the actor network.

\section{Performance Evaluation} \label{sec:performace-evaluation}
In this section, we first present the implementation of the thermal control simulation system, and then introduce the experiment settings and datasets, and compare the performances of our proposed methods with the baseline methods.
\subsection{Implementation of Thermal Control Simulation System} \label{sec:thermal-control-implementation}
\begin{figure}
\begin{center}
\epsfig{file=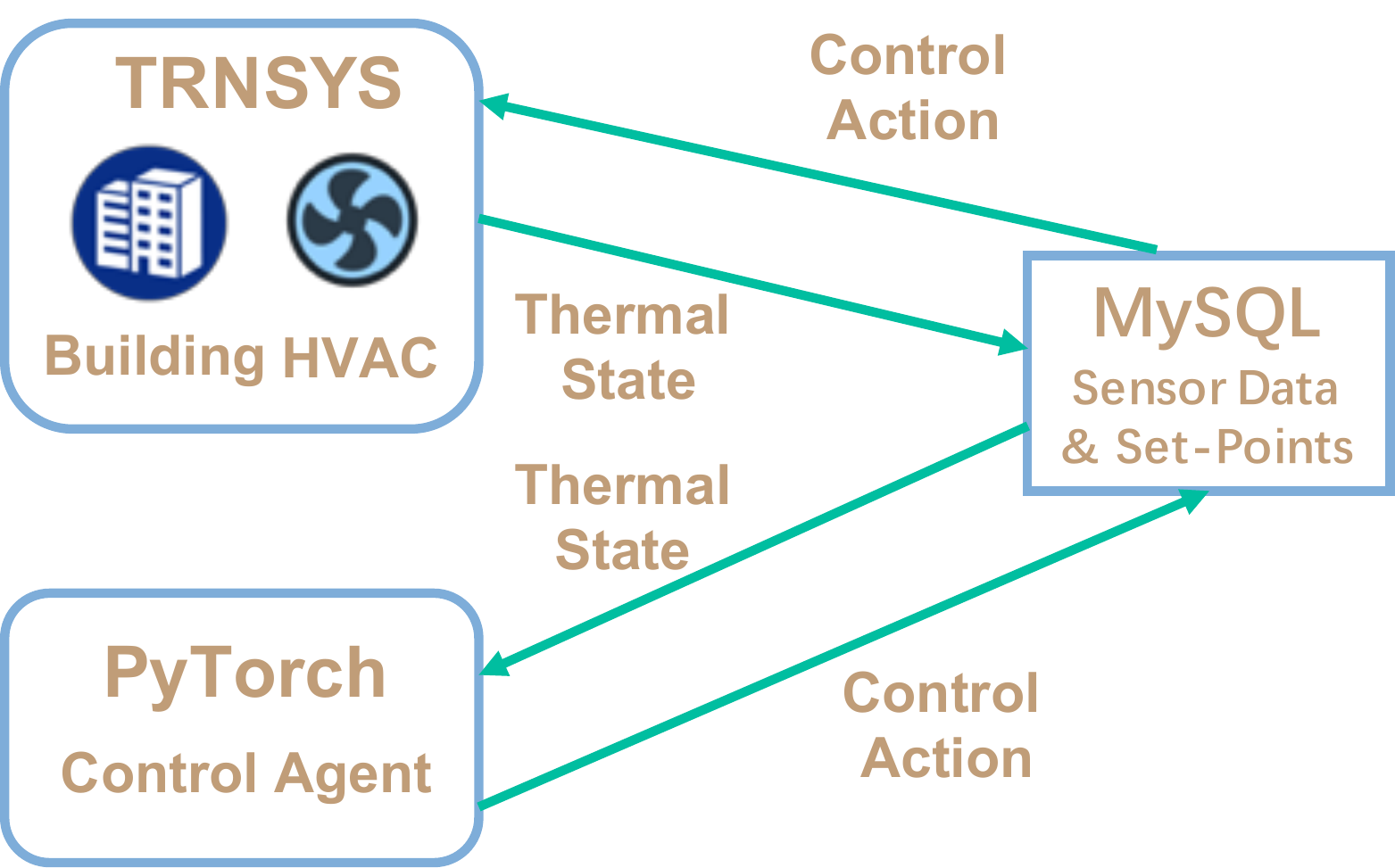, width=0.6\columnwidth}
\end{center}
\vspace{-3 mm}
\caption{The implementation of the thermal control simulation system. We adopt TRNSYS for simulating the building thermal environment and the HVAC system, MySQL for storing sensor data and the set-points of the HVAC system, PyTorch for implementing the control agent.} \label{fig:thermal-simulation}
\end{figure}
\begin{figure}
\begin{center}
\epsfig{file=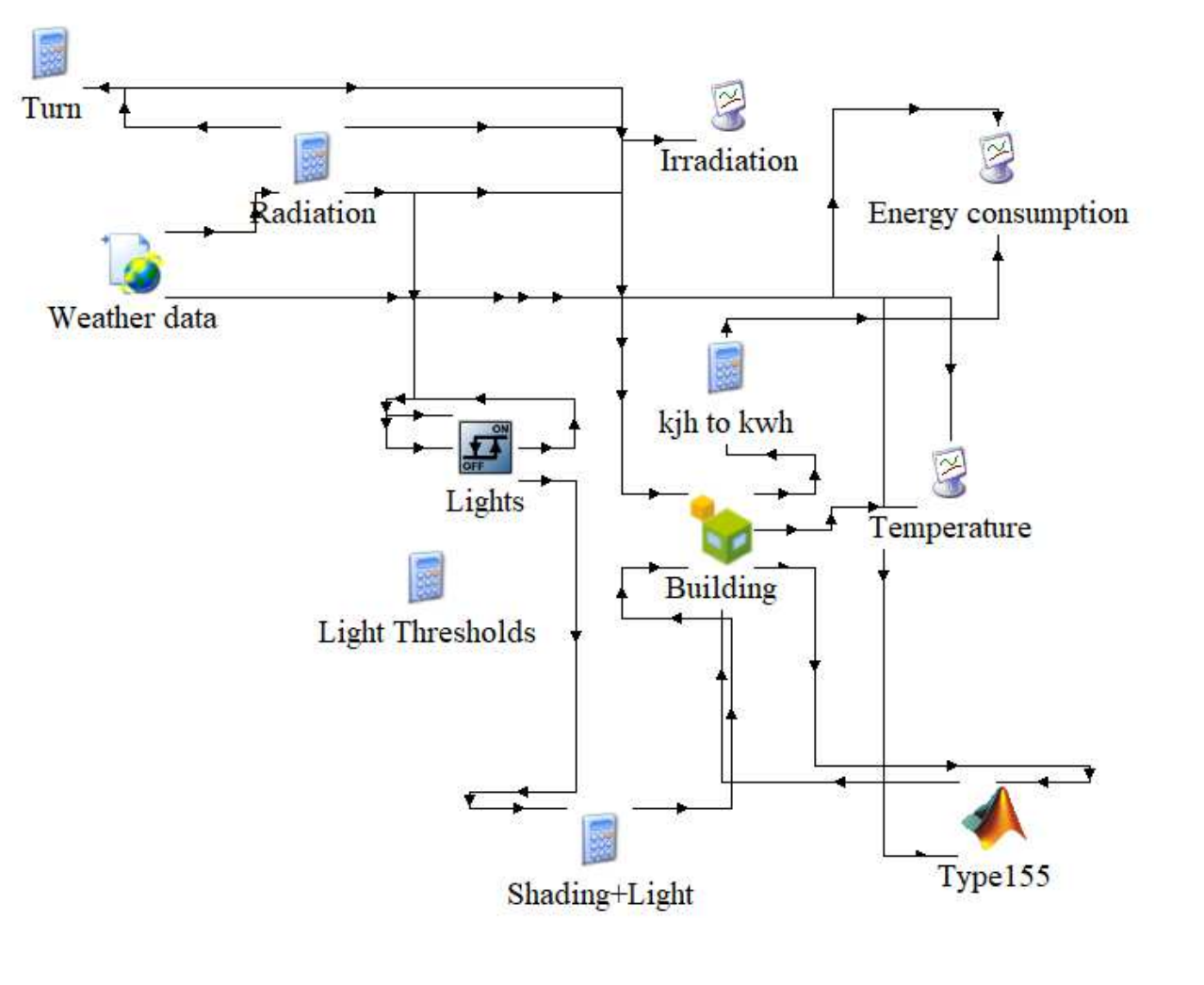, width=0.8\columnwidth}
\end{center}
\vspace{-3 mm}
\caption{The control diagram of building thermal simulation in TRNSYS.} \label{fig:trnsys-flow}
\end{figure}

We implement the thermal control simulation system for simulating the thermal dynamics in a building and the energy consumption of the equipped HVAC system.
The main components of the  thermal control simulation system is illustrated in Fig. \ref{fig:thermal-simulation}.
We use TRNSYS \cite{trnsys_url} to simulate the HVAC system and the thermal dynamics in the building.
The thermal control algorithm and the thermal comfort prediction algorithm are implemented with PyTorch \cite{pytorch_url}, which is an open-source machine learning library for Python.
We use MySQL database \cite{mysql_url} as an interface for the control interactions between the thermal control agent and TRNSYS,
because TRNSYS uses the Matlab-based programming interface for interacting with other programs.
The control diagram of the thermal simulation in TRNSYS is illustrated in Fig. \ref{fig:trnsys-flow}.

At the beginning of each time slot, the thermal state of the building will be read from TRNSYS and written into MySQL.
The control agent will read the thermal state information from MySQL and make a control action, which is the new set-points of the HVAC system.
The control action will be written into MySQL, and TRNSYS will read the control action from MySQL and update the set-points of the HVAC system.
TRNSYS will use the new set-points of the HVAC system and environment information (e.g., outside temperature, solar irradiation) to calculate the energy consumption of the HVAC system and the indoor temperature and humidity.
The control agent can improve the control policy during the iterations.
\subsection{Experiment Settings}
We simulate the building environment of a laboratory which is 307 $M^2$. 
It has 30 occupants and 40 computers with monitors. 
The heat gain from the occupants and computers will also influence the thermal conditions in the environment. 
The air change rate is 0.67/hour.
The weather dataset we use for our simulation is SG-Singapore-Airp-486980, which is collected from Singapore.
We use 10,000 hours of the simulation data in TRNSYS for training the models, and use 5,000 hours of data for testing the performances.
In our implementation of DDPG, the actor network and the critic network have two hidden layers, and each layer has 128 neurons.
We use the tanh activation function and batch normalization in each layer. 
We adopt Adam for gradient-based optimization and the learning rate is 0.001.
The discount factor $\tau$ for model update is 0.001 and the batch size is 128.
The duration of each time slot is 30 minutes and each episode consists of 48 time slots.
The default weight of energy cost is 0.05.
The initial exploration noise scale is 0.7 and the final noise scale is 0.1, and the exploration noise decreases linearly over 300 episodes to 0.1.

\subsection{Thermal Comfort Prediction Performance}
We adopt the ASHRAE RP-884 thermal comfort dataset \cite{de1998global} for training our deep neural network model for thermal comfort prediction.
We use 11164 samples from the dataset for training and testing, and each sample is a human subject's evaluation of the thermal comfort under a certain thermal condition.
The samples are randomly divided, and 80\% of the samples are used for training and 20\% of the samples are used for testing.
We compare the performance of our deep neural network based thermal prediction method with following baseline methods,
namely, Linear Regression, Support Vector Machine (SVM), Gaussian Process Regression, Ensemble Regression.
We evaluate the prediction errors of different methods using Mean Squared Error (MSE).

\begin{figure}
\begin{center}
\epsfig{file=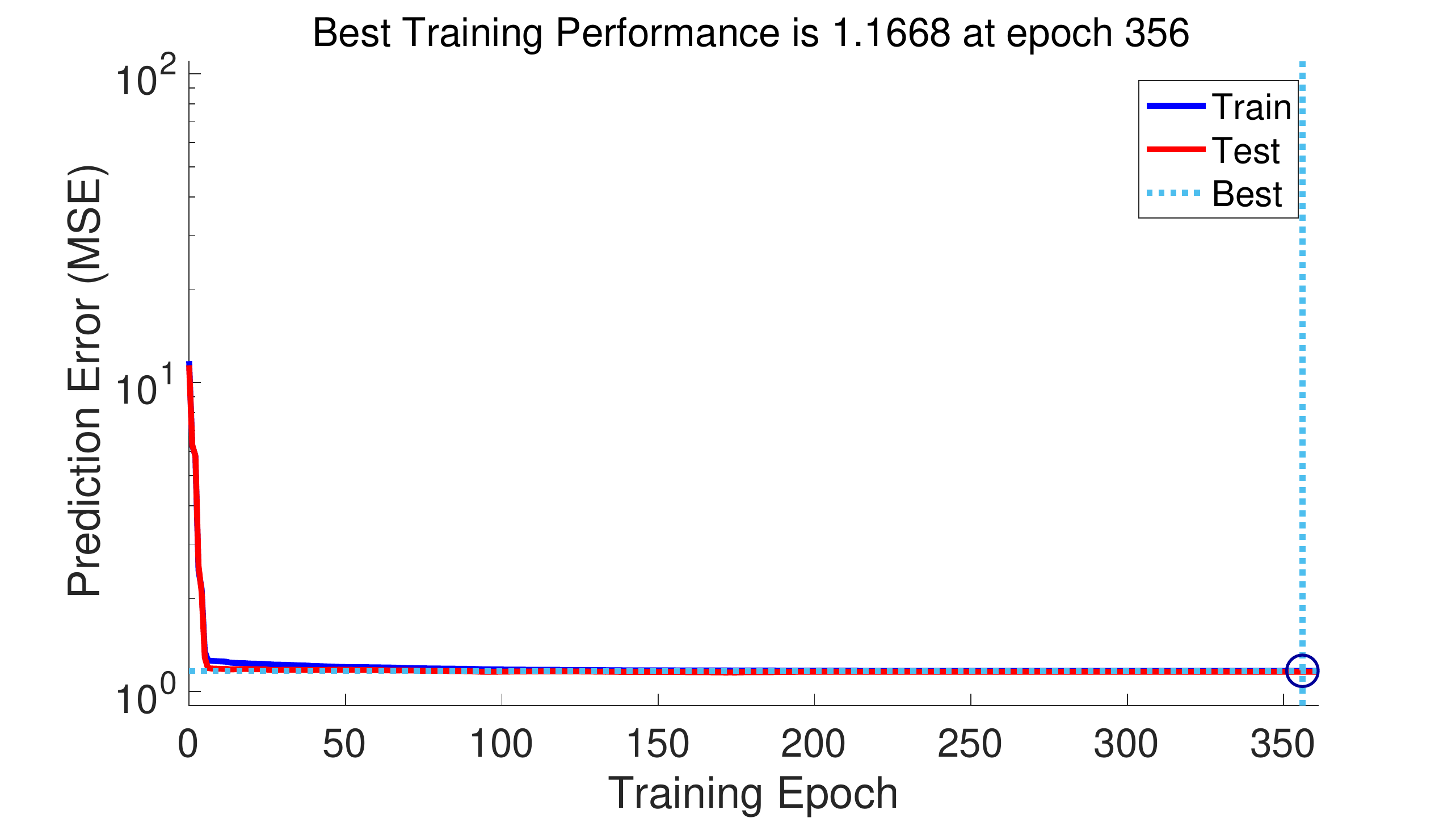, width=0.9\columnwidth}
\end{center}
\vspace{-3 mm}
\caption{The convergence of the deep neural network method for thermal comfort prediction. The prediction error can converge after several epochs of training, and the best training performance is 1.1668.} \label{fig:training-error-epoches}
\end{figure}

\begin{figure}
\begin{center}
\epsfig{file=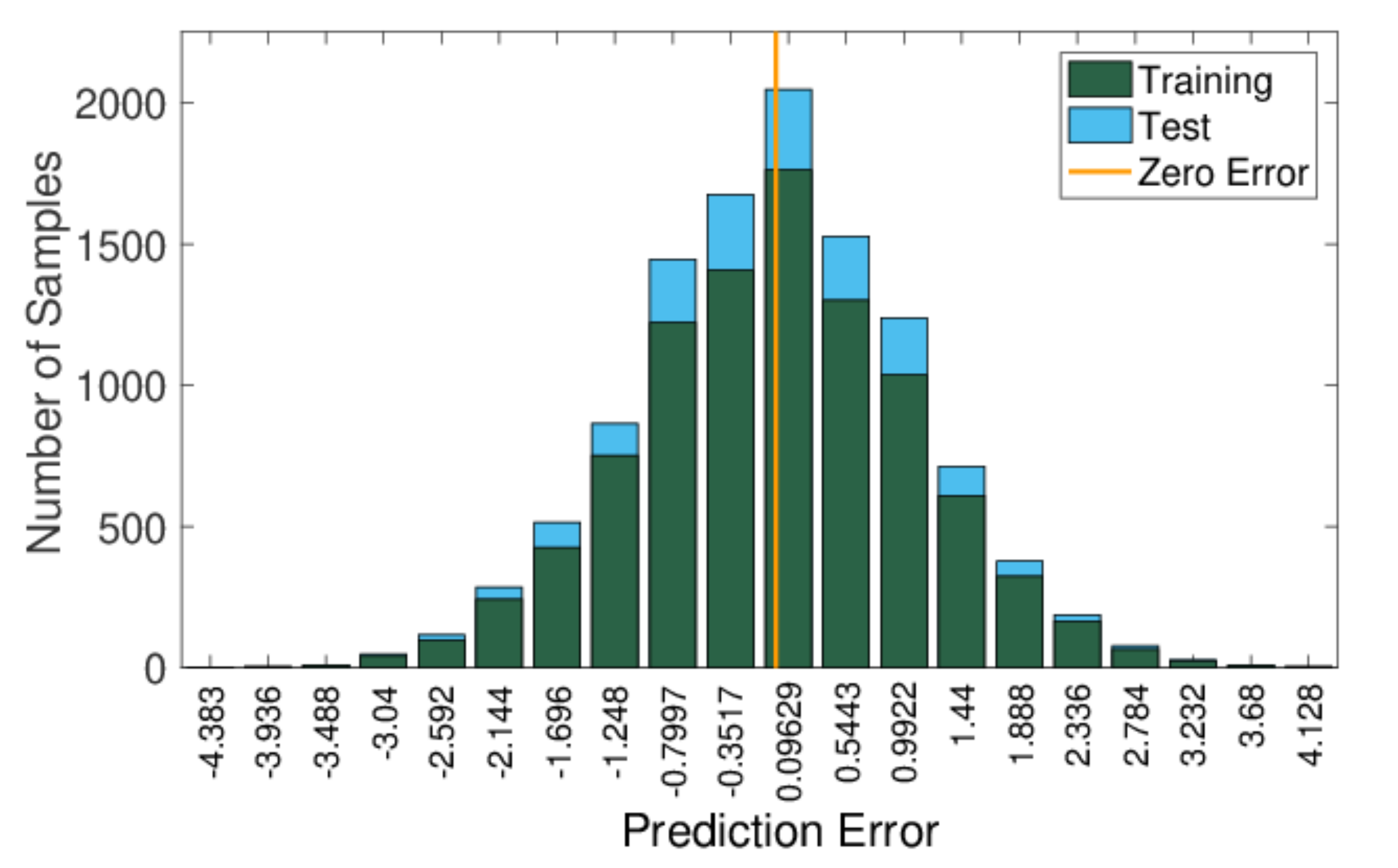, width=0.87\columnwidth}
\end{center}
\vspace{-3 mm}
\caption{The distribution of the prediction errors of the training and test samples. The prediction errors of most of the training samples and test samples are within the range [-1, 1].} \label{fig:prediction-error-dis}
\end{figure}

\begin{figure}
\begin{center}
\epsfig{file=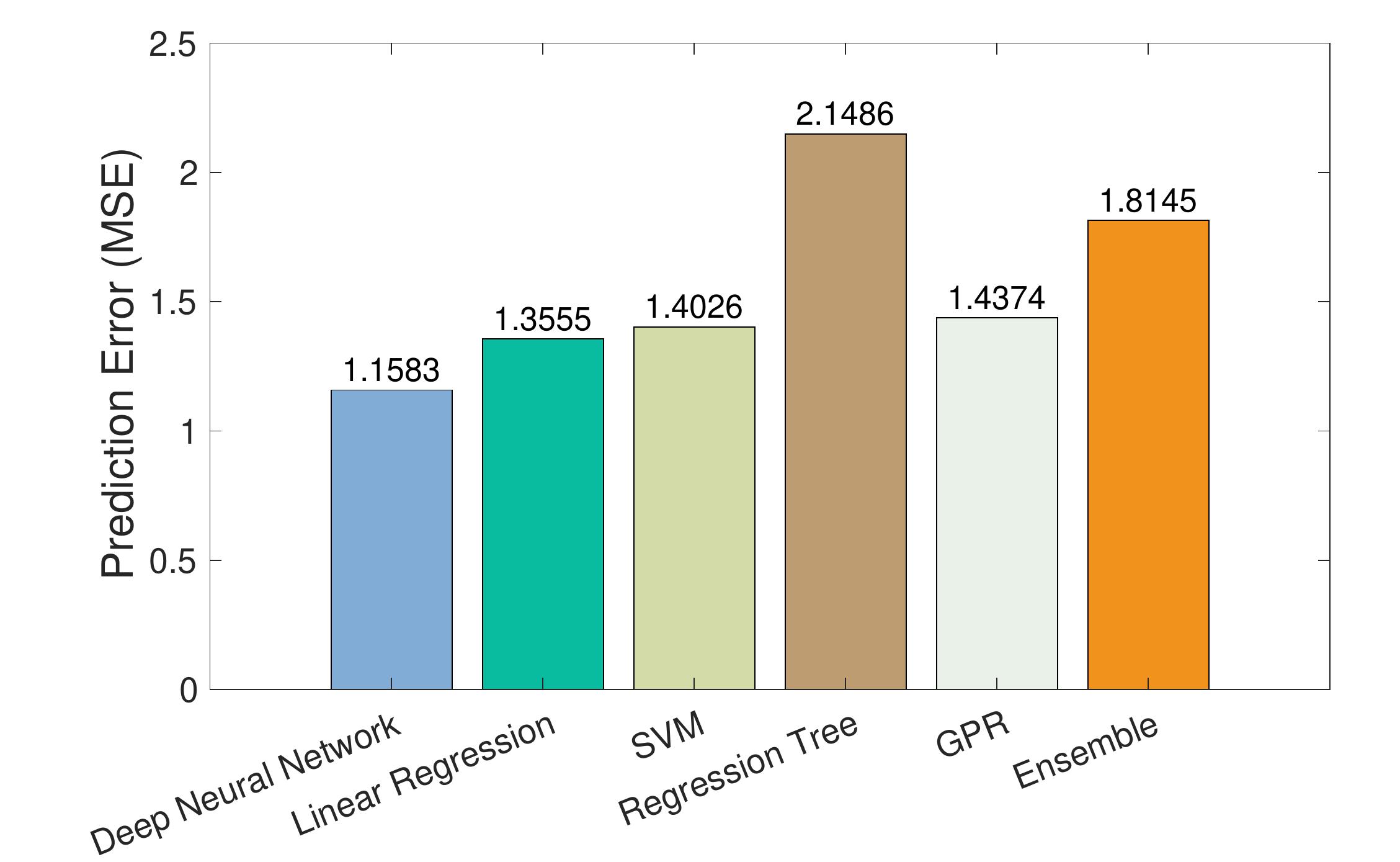, width=0.9\columnwidth}
\end{center}
\vspace{-3 mm}
\caption{The comparison of prediction errors of different methods. Our method can achieve smaller prediction error compared with the baseline methods.} \label{fig:thermal-comfort-prediction-cmp}
\end{figure}

We first illustrate the prediction error of our method during different training epochs in Fig. \ref{fig:training-error-epoches}.
The training error can converge after several epochs, and the best training performance is 1.1668 at epoch 356.
The prediction error distribution of our method is illustrated in Fig. \ref{fig:prediction-error-dis}. 
From the test, we can observe that for most of the training samples and test samples, the prediction error is within the range $[-1, 1]$.
The performance comparison of different methods is illustrated in Fig. \ref{fig:thermal-comfort-prediction-cmp}.
The MSE of our method is 1.1583, and the MSEs of Linear Regression, Support Vector Machine (SVM), Gaussian Process Regression, Ensemble Regression are 1.3555, 1.4026, 2.1486, 1.4374, 1.8145, respectively.
It can be observed that our method can achieve smaller prediction error compared with the baseline methods.
The results verify that our method can achieve a higher prediction accuracy, therefore, it can be adopted for predicting the occupants' thermal comfort more precisely compared  with the other baseline methods.

\begin{figure}
\begin{center}
\epsfig{file=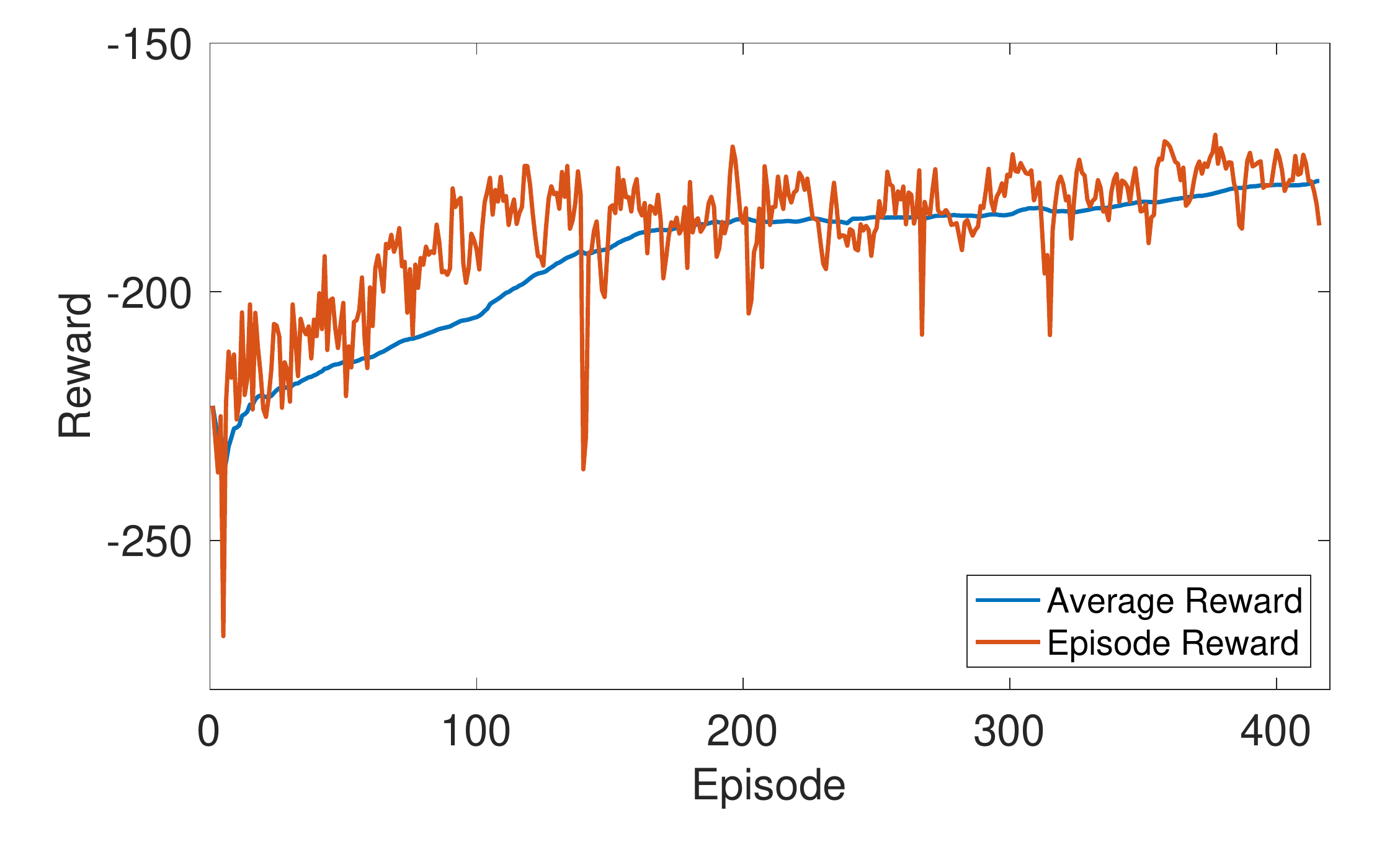, width=0.9\columnwidth}
\end{center}
\vspace{-3 mm}
\caption{The convergence of the DDPG based thermal control algorithm. The average reward improves during the training and converges.} \label{fig:ddpg-convergence-training}
\end{figure}

\subsection{Performance under Different Experiment Settings}
In this subsection, we evaluate the performances of the DDPG based thermal control method under different settings. 

\emph{Convergence of the DDPG based thermal control.}
We illustrate the rewards of the DDPG based thermal control method during different training episodes in Fig. \ref{fig:ddpg-convergence-training}.
The DDPG algorithm updates the policy during training, and we can observe that the received reward during each episode improves over the training stage and finally converges. 
As illustrated in Fig. \ref{fig:ddpg-convergence-training}, the reward received during each episode (Episode Reward) fluctuates during the training, 
this is because the exploration noise and the varying ambient thermal conditions in each episode affect the energy consumption and thermal comfort,
leading to the fluctuant rewards in each episode.
The average reward illustrated in Fig. \ref{fig:ddpg-convergence-training} is the averaged reward of the past 100 episodes and reflects the changing trend of the reward.
We can observe that the average reward improves stably during the training and converges.

\begin{figure}
\begin{center}
\epsfig{file=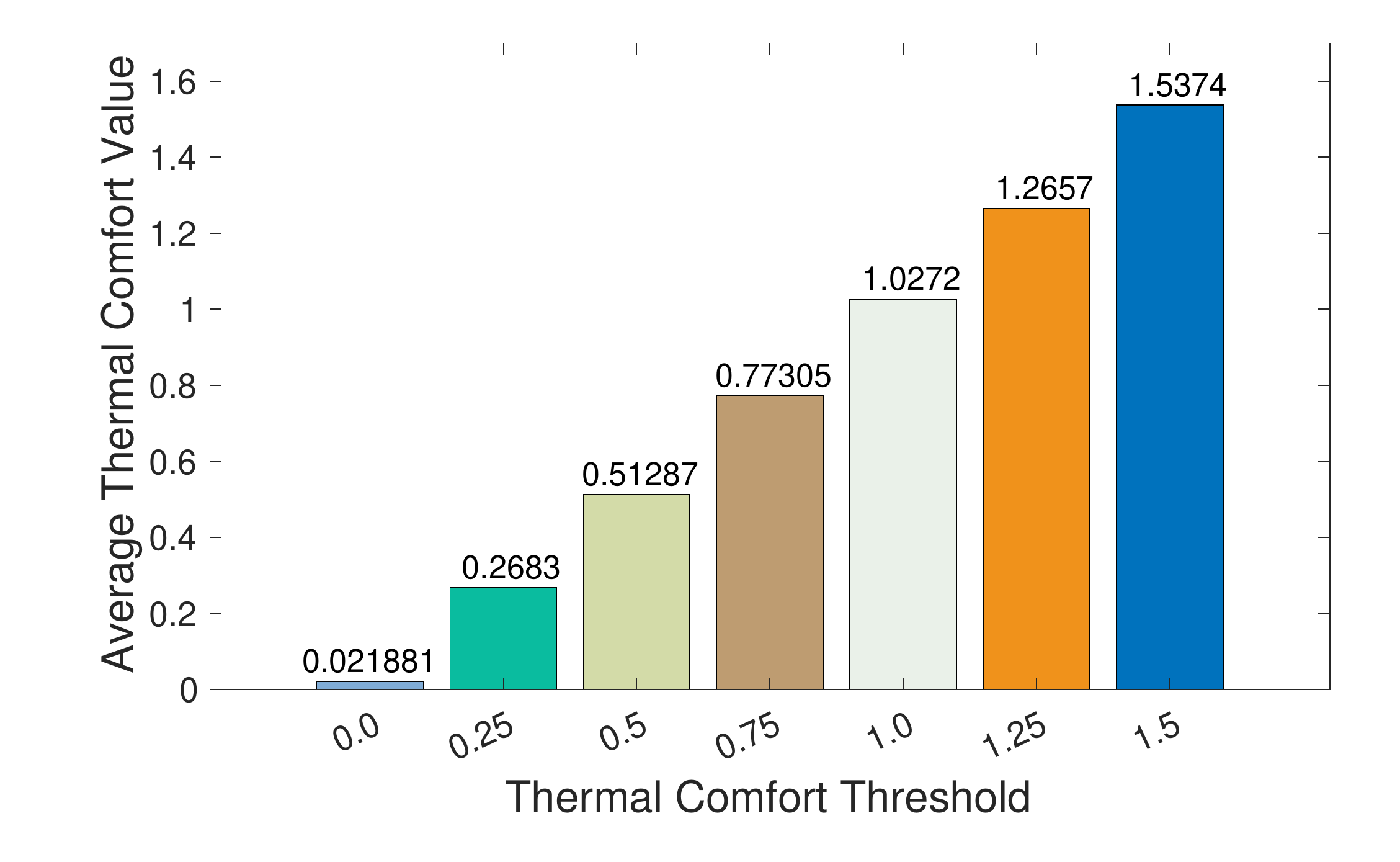, width=0.9\columnwidth}
\end{center}
\vspace{-3 mm}
\caption{The average thermal comfort values under different thermal comfort thresholds. The average thermal comfort values are close to the thresholds.} \label{fig:pmv-under-different-threshoolds}
\end{figure}

\begin{figure}
\begin{center}
\epsfig{file=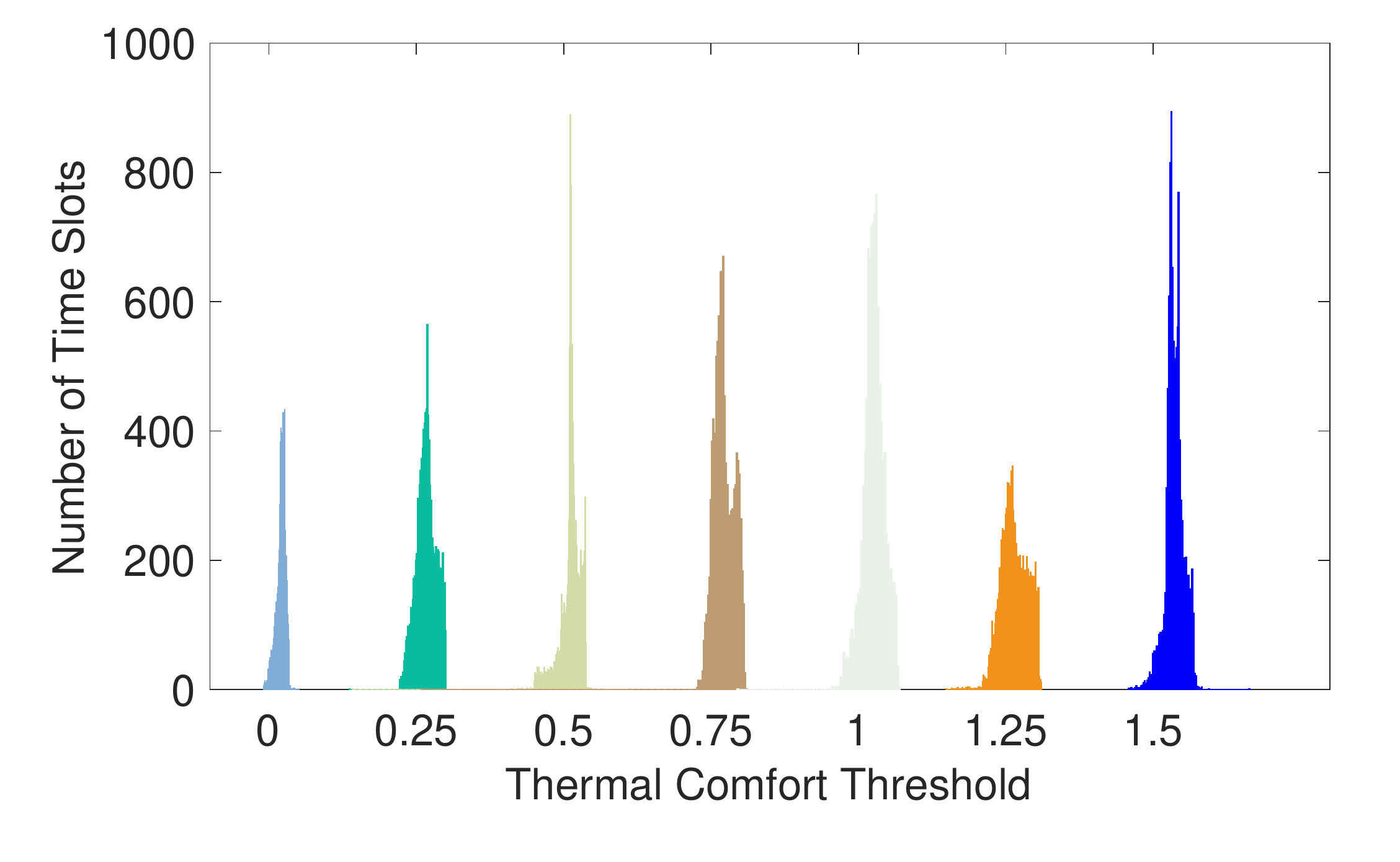, width=0.9\columnwidth}
\end{center}
\vspace{-3 mm}
\caption{The distribution of the thermal comfort values under different thermal comfort thresholds. The thermal comfort value of each time slot is closely centered around the prescribed thermal comfort threshold.} \label{fig:pmv-dis-different-thresholds}
\end{figure}

\emph{Performance under different thermal comfort thresholds.}
With our method one can set different thermal comfort thresholds according to the occupants' thermal comfort requirements.
We evaluate the energy consumption of the HVAC system and the distribution of the thermal comfort values under different thermal comfort thresholds.
In Fig. \ref{fig:pmv-under-different-threshoolds} we illustrate the average thermal comfort values under different prescribed thermal comfort thresholds,
and it can be observed that the  average actual thermal comfort value measured from the building is close to the prescribed thresholds.
In Fig. \ref{fig:pmv-dis-different-thresholds} we illustrate the distribution of the thermal comfort value of each time slot under different prescribed thermal comfort thresholds.
It can be verified that the thermal comfort values in the indoor environment are closely centered around the prescribed thresholds.
Therefore, our method can control the indoor thermal comfort precisely, given the occupants' thermal comfort requirements and appropriate settings.

\begin{figure}
\begin{center}
\epsfig{file=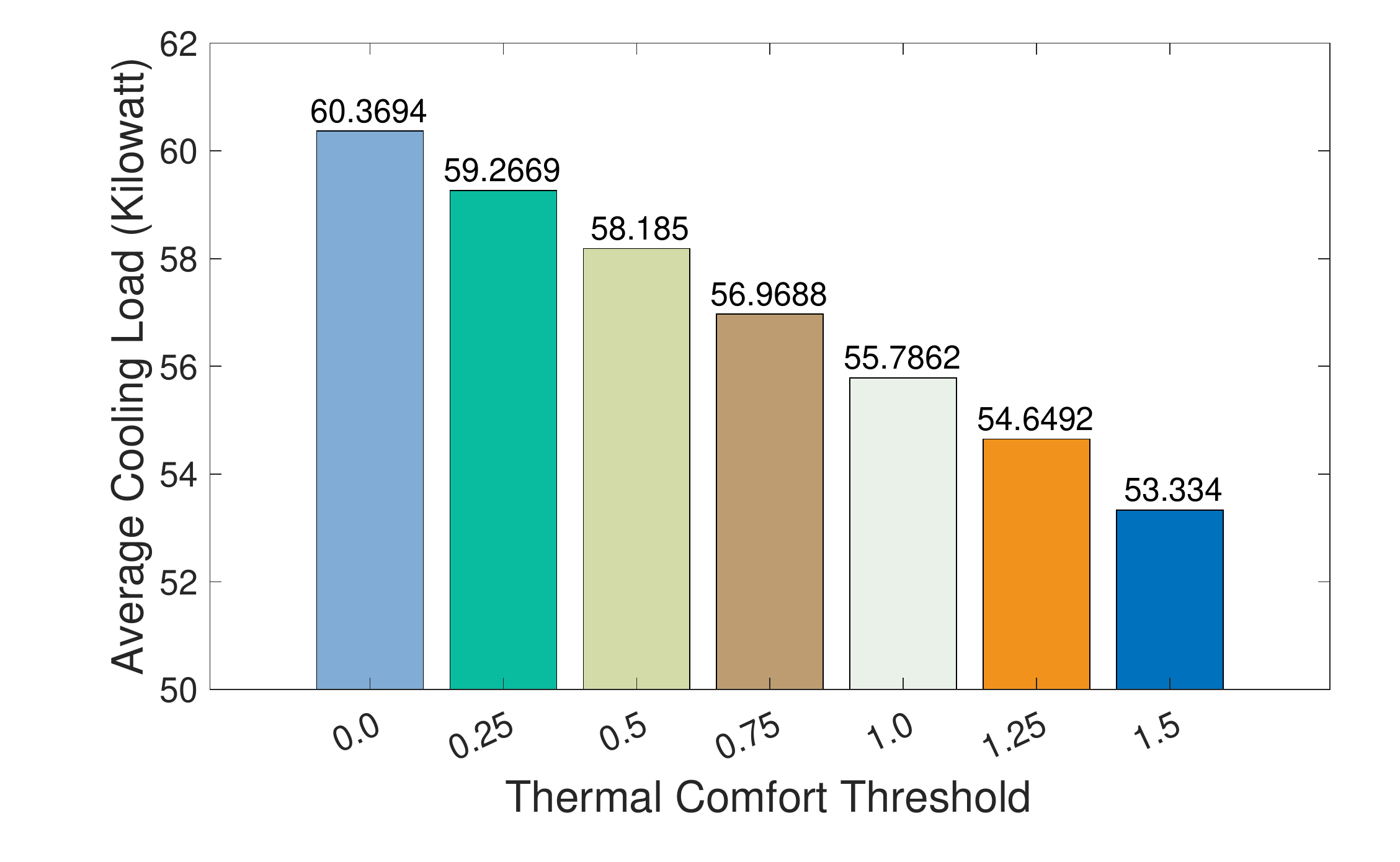, width=0.9\columnwidth}
\end{center}
\vspace{-3 mm}
\caption{The average cooling load of the HVAC system under different thermal comfort thresholds. A larger thermal comfort threshold will result in less energy consumption of the HVAC system.} \label{fig:energy-consumption-under-different-thresholds}
\end{figure}

\begin{figure}
\begin{center}
\epsfig{file=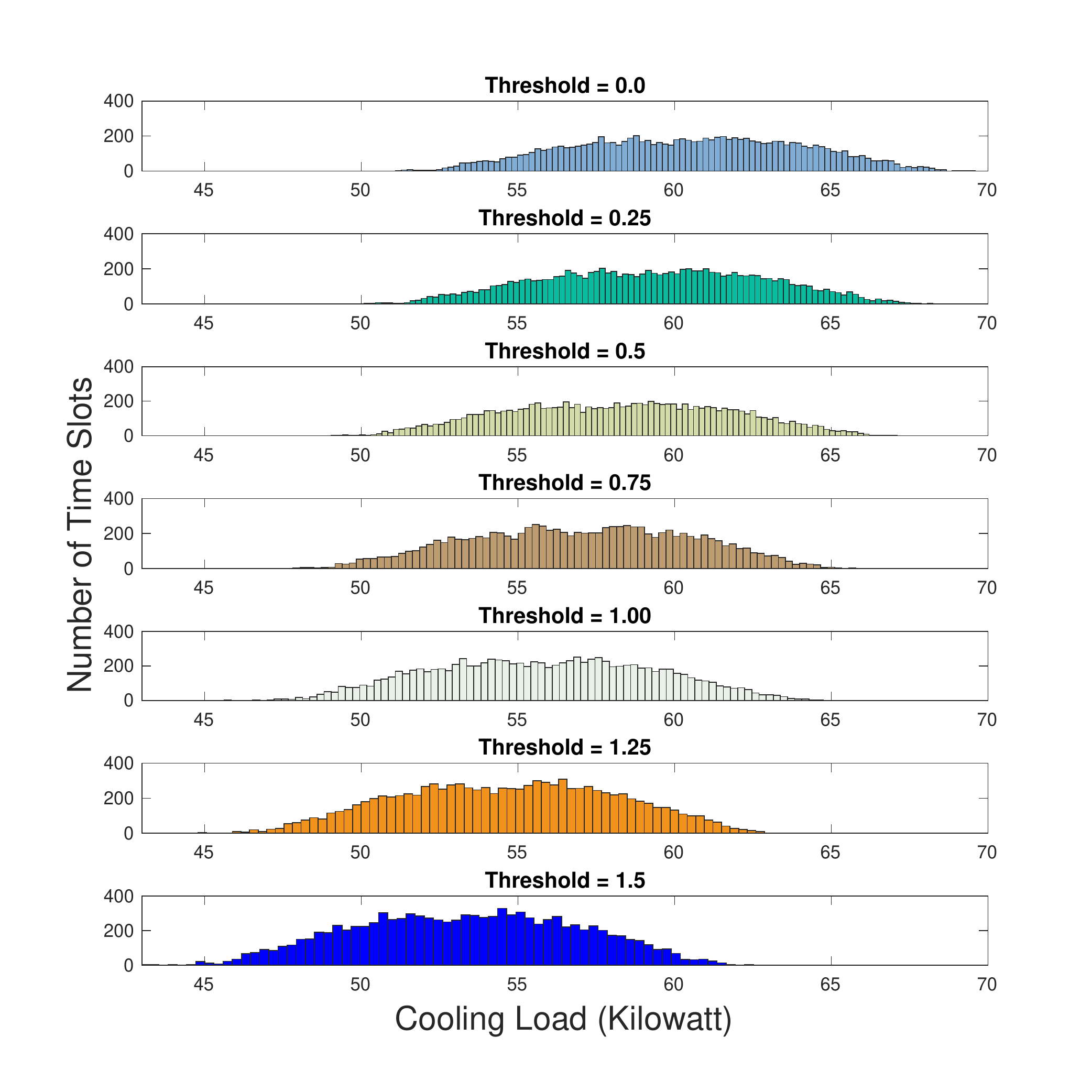, width=1.0\columnwidth}
\end{center}
\vspace{-3 mm}
\caption{The distribution of the cooling load of the HVAC system under different thermal comfort thresholds. The distribution of cooling load moves to smaller values under a larger thermal comfort threshold.} 
\label{fig:energy-distribution-different-thresholds}
\end{figure}

In Fig. \ref{fig:energy-consumption-under-different-thresholds} we illustrate the average cooling load of the HVAC system under different thermal comfort thresholds.
It will lead to less energy consumption of the HVAC system if the thermal comfort threshold is set to a larger value.
Therefore, if the occupants do not have stringent thermal comfort requirement, the thermal comfort threshold can be set to a larger value for energy efficiency.
We illustrate the distribution of the cooling load of the HVAC system under different thermal comfort thresholds in Fig. \ref{fig:energy-distribution-different-thresholds}.
We can observe that the distribution of the cooling load move towards small values when the thermal comfort threshold is larger.
Therefore, adjusting the thermal comfort threshold can control the energy consumption of the HVAC system.

\begin{figure}
\begin{center}
\epsfig{file=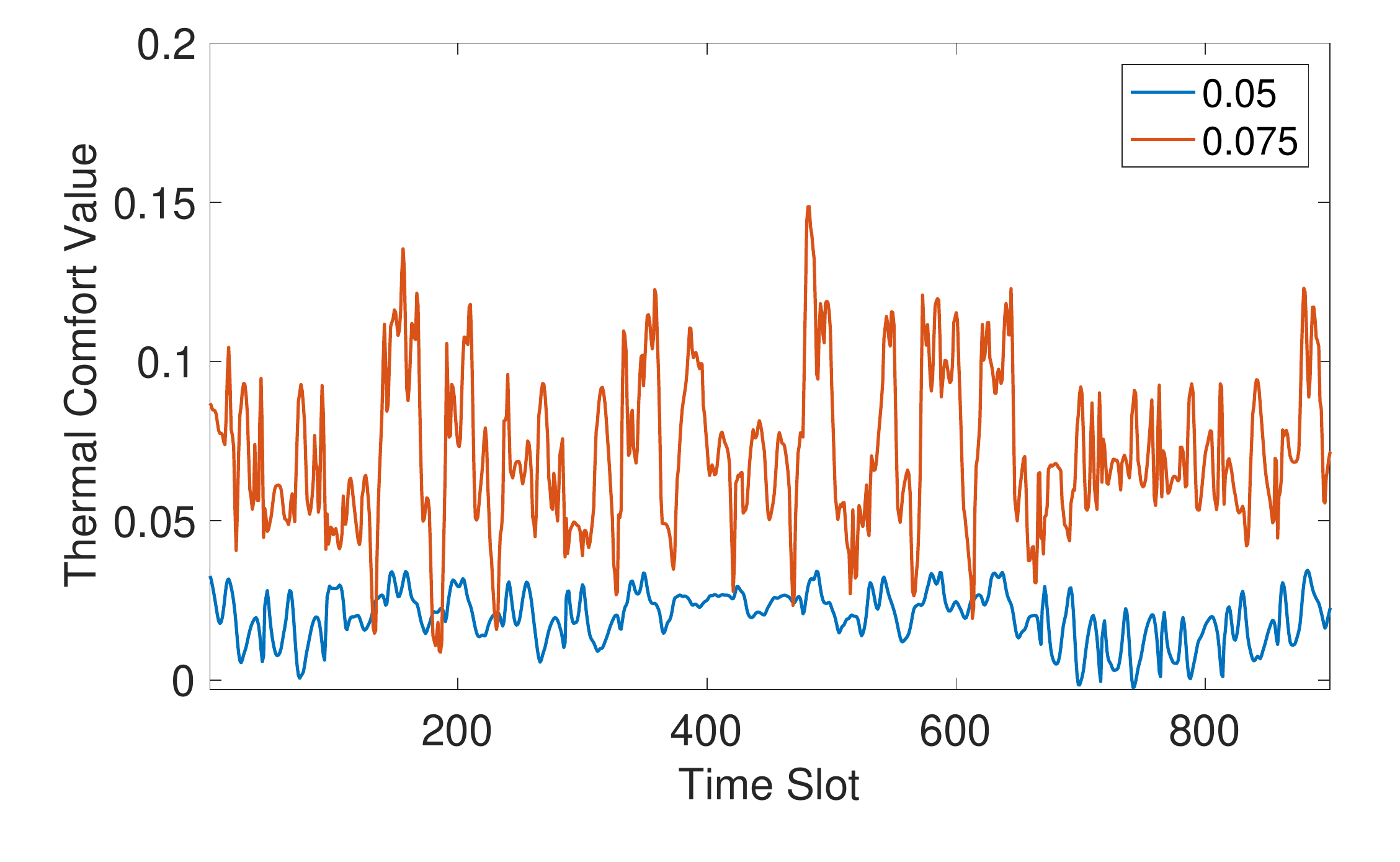, width=0.95\columnwidth}
\end{center}
\vspace{-3 mm}
\caption{The thermal comfort values under different weights of energy cost over time. If the weight of the energy cost is small, the thermal comfort value will be close to the threshold, and the changes of the value will be smaller.} \label{fig:pmv-under-different-weight}
\end{figure}

\begin{figure}
\begin{center}
\epsfig{file=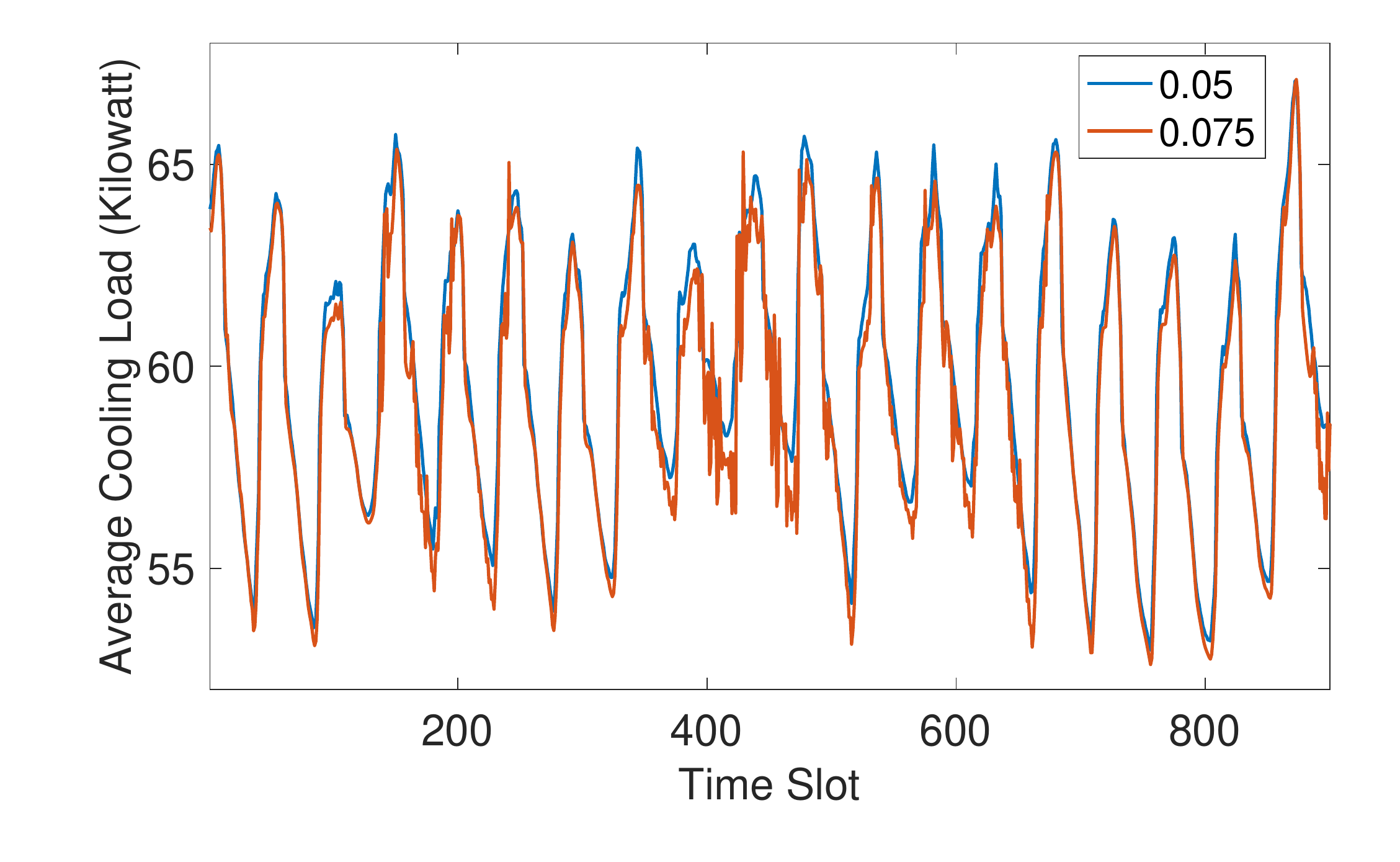, width=0.95\columnwidth}
\end{center}
\vspace{-3 mm}
\caption{The time-varying cooling load of the HVAC system under different weights of energy cost. It will incur less energy consumption if the weight of energy cost is set as a larger value.} \label{fig:energy-consumption-different-weight}
\end{figure}

\emph{Performance under different weights of energy cost.}
We can set different weights for the cost of the energy consumption in the reward function (Eq. \ref{eqn:reward-function}).
If one puts more interests on the occupants' thermal comfort, the weight can be set smaller.
On the contrary, if one puts more interests on energy cost, the weight can be set larger.
We evaluate the performances of our method under different weights, and the results are shown in Fig. \ref{fig:pmv-under-different-weight} and \ref{fig:energy-consumption-different-weight}.
In Fig. \ref{fig:pmv-under-different-weight} we illustrate the changes of the thermal comfort value over time under different weights of energy cost.
The thermal comfort threshold in the experiment is 0.0.
We can observe that if the weight is small, the thermal comfort value will be close to the threshold, and the changes of the thermal comfort value will be small.
This is because the penalty for violating the thermal comfort threshold is larger than the energy cost, due to the smaller weight of the energy cost.
Therefore, the thermal comfort value will be kept close to the threshold for reducing the cost incurred by violating the thermal comfort threshold.
On the contrary, if the weight of the energy cost is large, the thermal comfort value may fluctuate largely for reducing the energy consumption.
We illustrate the cooling load of the HVAC system over time under different weights of energy cost in Fig. \ref{fig:energy-consumption-different-weight}.
It can be observed that it will incur less energy consumption if the weight of the energy cost is high.
This is because it will incur more cost if the HVAC system keeps maintaining the thermal comfort value within the threshold, which will incur high energy consumption.

\subsection{Performance Comparison with Different Methods}
\begin{figure}
\begin{center}
\epsfig{file=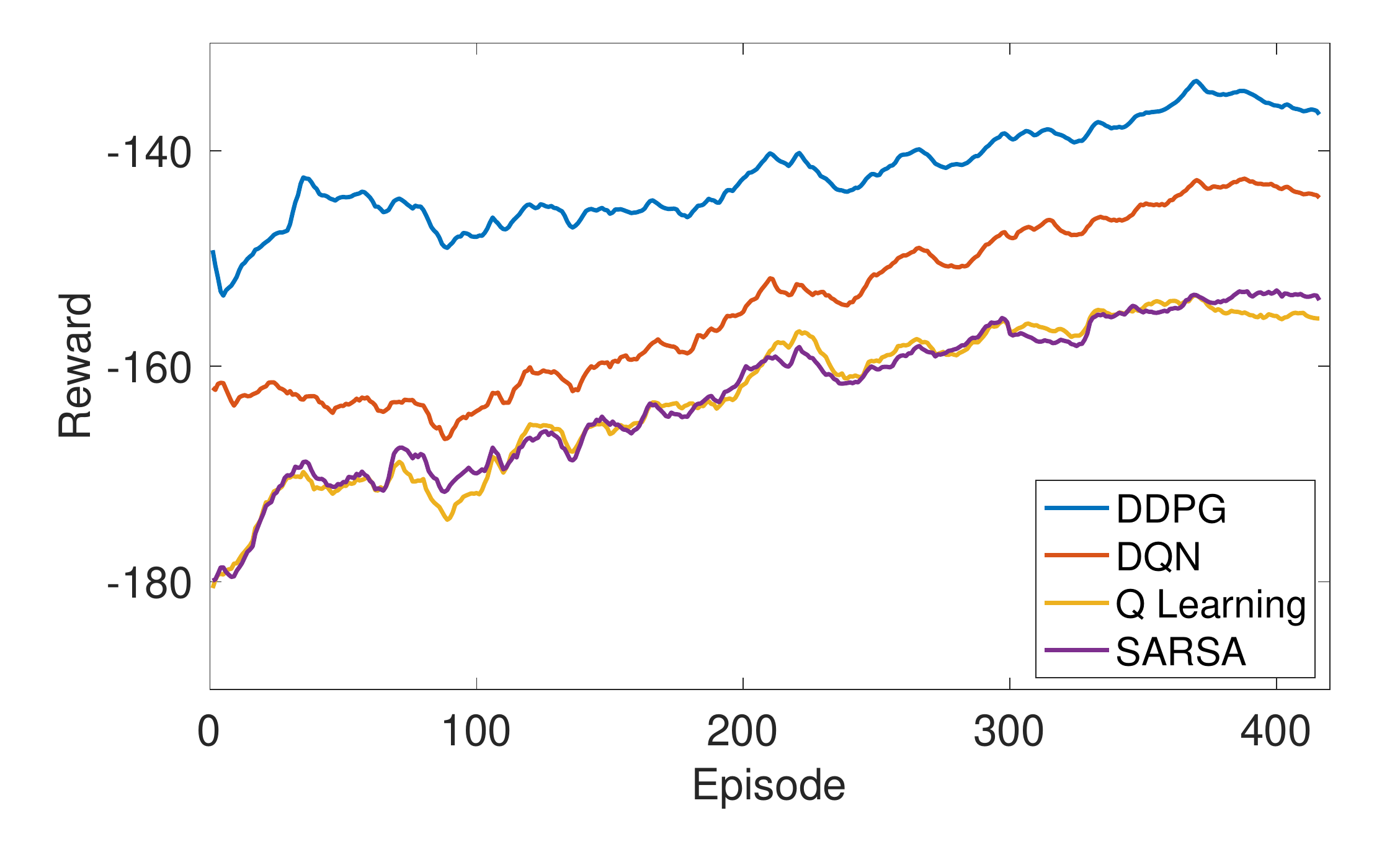, width=0.85\columnwidth}
\end{center}
\vspace{-3 mm}
\caption{The convergence of different algorithms. DDPG can converge faster and achieve a higher reward than the other baseline methods.} 
\label{fig:cmp-convergence}
\end{figure}

\begin{figure}
\begin{center}
\epsfig{file=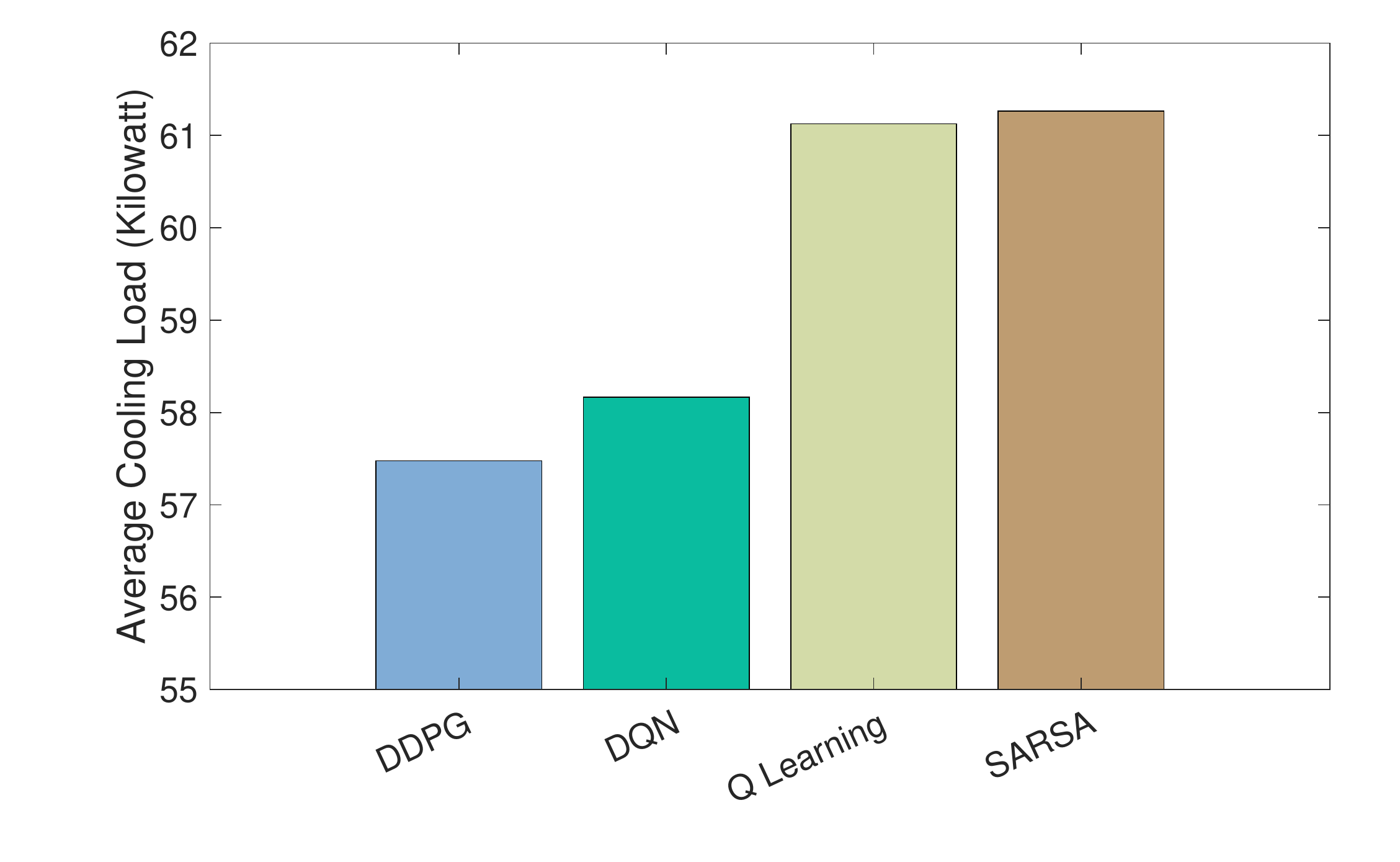, width=0.85\columnwidth}
\end{center}
\vspace{-3 mm}
\caption{The average cooling load of different methods. DDPG can achieve less energy consumption compared with the baseline methods.} 
\label{fig:cmp-energy}
\end{figure}

\begin{figure}
\begin{center}
\epsfig{file=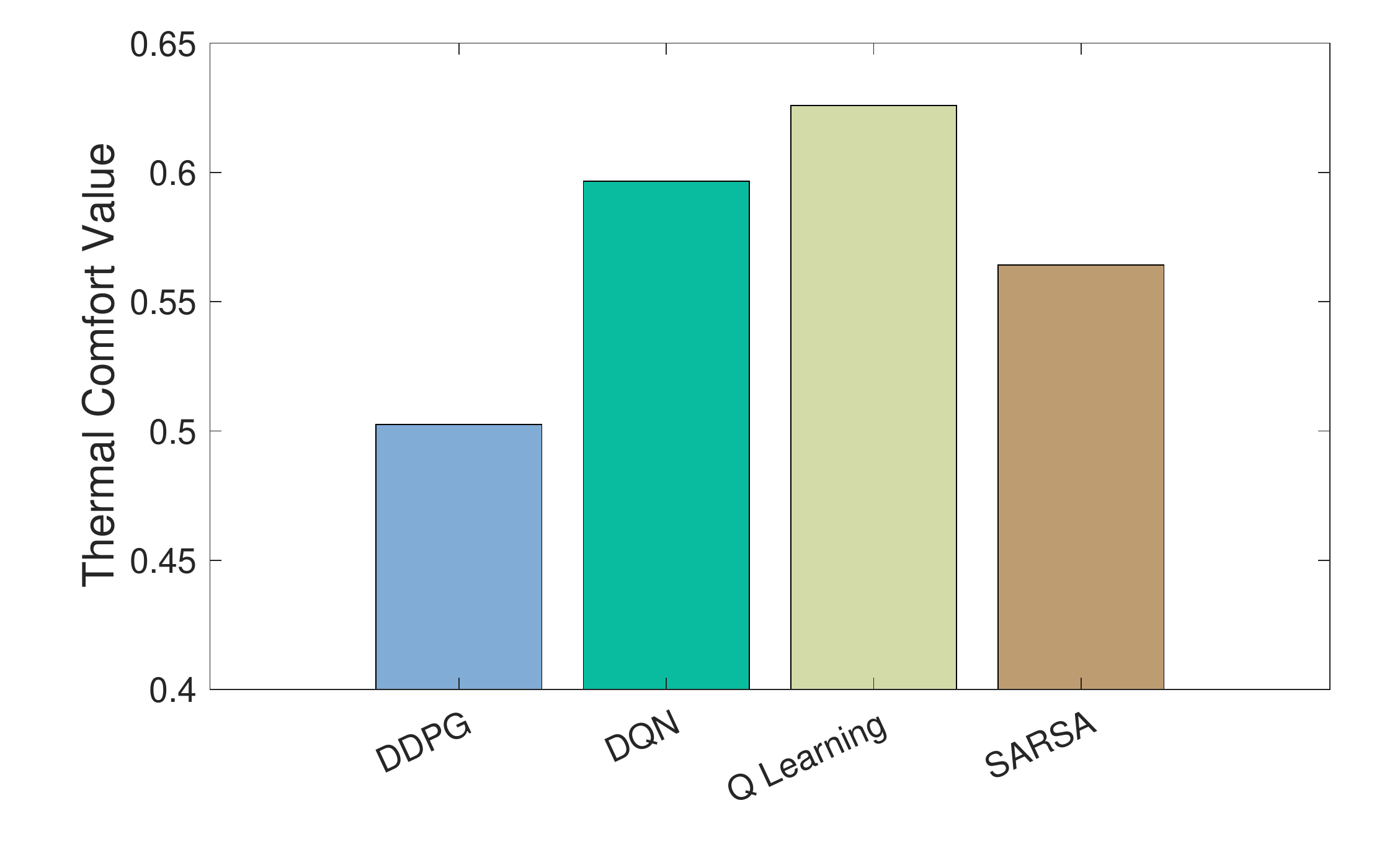, width=0.85\columnwidth}
\end{center}
\vspace{-3 mm}
\caption{The average thermal comfort values of different methods. DDPG can achieve higher thermal comfort (i.e., smaller thermal comfort value) compared with baseline methods.} \label{fig:cmp-thermal}
\end{figure}

We compare the performance of our thermal control method (DDPG) with the following baseline methods, namely, Q Learning, SARSA, and DQN.
In the baseline methods, the temperature is discretized with the granularity of one centi-degree and the humidity is discretized with the granularity of 5 percentages. 
We illustrate the convergences of different methods in Fig. \ref{fig:cmp-convergence}.
It can be observed that DDPG can achieve a faster convergence speed compared with the other baseline methods.
This is because DDPG does not need the discretization of the action space and has fewer number of outputs of the network.
Therefore, DDPG can learn for thermal control more efficiently. 
Moreover, DDPG can also achieve a higher reward compared with the baseline methods.
This verifies that our method can achieve higher performances compared with the baseline methods.
This is because Q learning and SARSA adopt the tabular methods for storing and updating the state-action values without generalization,
and DQN needs the discretization of the action space.
We illustrate the average cooling load of different methods in Fig. \ref{fig:cmp-energy}.
It can be observed that the average cooling load of our method is lower than the other baseline methods,
therefore, our method can achieve higher energy-efficiency. 
We illustrate the average thermal comfort values of different methods in Fig. \ref{fig:cmp-thermal}. 
The average thermal comfort value of DDPG is more close to the preset threshold 0.5 and lower than the baseline methods.
Therefore, our method can control the set-points of the HVAC system more precisely and it can achieve a higher degree of thermal comfort compared with the other methods.

\section{Conclusion} \label{sec:conclusion}
In this paper, we proposed a learning based optimization framework for optimizing the occupants' thermal comfort and the energy consumption of the HVAC systems in smart buildings.
We first designed a deep neural network based method with Bayesian regularization for thermal comfort prediction,
and then we adopted DDPG for controlling the HVAC system for optimizing the energy consumption whiling meeting the occupants' thermal comfort requirements.
For verifying the effectiveness of our proposed methods, we implement a building thermal control simulation system using TRNSYS.
We evaluated the performances of our methods under different settings.
The results show that our method can achieve better prediction performances for thermal comfort prediction,
and it can achieve higher thermal comfort and energy-efficiency compared with the baseline methods.
In future works, we may consider using transfer learning to improve learning efficiency via the transfer of knowledge from different HVAC systems.

\bibliographystyle{IEEEtran}
\bibliography{thermal}

\begin{thebibliography}{10}
\providecommand{\url}[1]{#1}
\csname url@samestyle\endcsname
\providecommand{\newblock}{\relax}
\providecommand{\bibinfo}[2]{#2}
\providecommand{\BIBentrySTDinterwordspacing}{\spaceskip=0pt\relax}
\providecommand{\BIBentryALTinterwordstretchfactor}{4}
\providecommand{\BIBentryALTinterwordspacing}{\spaceskip=\fontdimen2\font plus
\BIBentryALTinterwordstretchfactor\fontdimen3\font minus
  \fontdimen4\font\relax}
\providecommand{\BIBforeignlanguage}[2]{{%
\expandafter\ifx\csname l@#1\endcsname\relax
\typeout{** WARNING: IEEEtran.bst: No hyphenation pattern has been}%
\typeout{** loaded for the language `#1'. Using the pattern for}%
\typeout{** the default language instead.}%
\else
\language=\csname l@#1\endcsname
\fi
#2}}
\providecommand{\BIBdecl}{\relax}
\BIBdecl

\bibitem{perez2008review}
L.~P{\'e}rez-Lombard, J.~Ortiz, and C.~Pout, ``A review on buildings energy
  consumption information,'' \emph{Energy and buildings}, vol.~40, no.~3, pp.
  394--398, 2008.

\bibitem{corgnati2007perception}
S.~P. Corgnati, M.~Filippi, and S.~Viazzo, ``Perception of the thermal
  environment in high school and university classrooms: Subjective preferences
  and thermal comfort,'' \emph{Building and environment}, vol.~42, no.~2, pp.
  951--959, 2007.

\bibitem{dounis2009advanced}
A.~I. Dounis and C.~Caraiscos, ``Advanced control systems engineering for
  energy and comfort management in a building environment—{A} review,''
  \emph{Renewable and Sustainable Energy Reviews}, vol.~13, no. 6-7, pp.
  1246--1261, 2009.

\bibitem{levermore1992building}
G.~J. Levermore, \emph{Building energy management systems: An application to
  heating and control}.\hskip 1em plus 0.5em minus 0.4em\relax E \& FN Spon,
  1992.

\bibitem{dounis1996comparison}
A.~I. Dounis, M.~Bruant, M.~Santamouris, G.~Guarracino, and P.~Michel,
  ``Comparison of conventional and fuzzy control of indoor air quality in
  buildings,'' \emph{Journal of Intelligent \& Fuzzy Systems}, vol.~4, no.~2,
  pp. 131--140, 1996.

\bibitem{shepherd2003fuzzy}
A.~Shepherd and W.~Batty, ``Fuzzy control strategies to provide cost and energy
  efficient high quality indoor environments in buildings with high occupant
  densities,'' \emph{Building Services Engineering Research and Technology},
  vol.~24, no.~1, pp. 35--45, 2003.

\bibitem{calvino2004control}
F.~Calvino, M.~La~Gennusa, G.~Rizzo, and G.~Scaccianoce, ``The control of
  indoor thermal comfort conditions: Introducing a fuzzy adaptive controller,''
  \emph{Energy and buildings}, vol.~36, no.~2, pp. 97--102, 2004.

\bibitem{maasoumy2011model}
M.~Maasoumy, A.~Pinto, and A.~Sangiovanni-Vincentelli, ``Model-based
  hierarchical optimal control design for {HVAC} systems,'' in \emph{ASME 2011
  Dynamic Systems and Control Conference and Bath/ASME Symposium on Fluid Power
  and Motion Control}.\hskip 1em plus 0.5em minus 0.4em\relax American Society
  of Mechanical Engineers, 2011, pp. 271--278.

\bibitem{barrett2015autonomous}
E.~Barrett and S.~Linder, ``Autonomous {HVAC} control, a reinforcement learning
  approach,'' in \emph{Joint European Conference on Machine Learning and
  Knowledge Discovery in Databases}.\hskip 1em plus 0.5em minus 0.4em\relax
  Springer, 2015, pp. 3--19.

\bibitem{li2015multi}
B.~Li and L.~Xia, ``A multi-grid reinforcement learning method for energy
  conservation and comfort of {HVAC} in buildings,'' in \emph{Automation
  Science and Engineering (CASE), 2015 IEEE International Conference on}.\hskip
  1em plus 0.5em minus 0.4em\relax IEEE, 2015, pp. 444--449.

\bibitem{nikovski2013method}
D.~Nikovski, J.~Xu, and M.~Nonaka, ``A method for computing optimal set-point
  schedules for {HVAC} systems,'' in \emph{REHVA World Congress CLIMA'13},
  2013.

\bibitem{zenger2013towards}
A.~Zenger, J.~Schmidt, and M.~Kr{\"o}del, ``Towards the intelligent home: Using
  reinforcement-learning for optimal heating control,'' in \emph{Annual
  Conference on Artificial Intelligence}.\hskip 1em plus 0.5em minus
  0.4em\relax Springer, 2013, pp. 304--307.

\bibitem{fazenda2014using}
P.~Fazenda, K.~Veeramachaneni, P.~Lima, and U.-M. O'Reilly, ``Using
  reinforcement learning to optimize occupant comfort and energy usage in
  {HVAC} systems,'' \emph{Journal of Ambient Intelligence and Smart
  Environments}, vol.~6, no.~6, pp. 675--690, 2014.

\bibitem{dalamagkidis2007reinforcement}
K.~Dalamagkidis, D.~Kolokotsa, K.~Kalaitzakis, and G.~S. Stavrakakis,
  ``Reinforcement learning for energy conservation and comfort in buildings,''
  \emph{Building and environment}, vol.~42, no.~7, pp. 2686--2698, 2007.

\bibitem{wei2017deep}
T.~Wei, Y.~Wang, and Q.~Zhu, ``Deep reinforcement learning for building {HVAC}
  control,'' in \emph{Design Automation Conference (DAC), 2017 54th
  ACM/EDAC/IEEE}.\hskip 1em plus 0.5em minus 0.4em\relax IEEE, 2017, pp. 1--6.

\bibitem{wang2017long}
Y.~Wang, K.~Velswamy, and B.~Huang, ``A long-short term memory recurrent neural
  network based reinforcement learning controller for office heating
  ventilation and air conditioning systems,'' \emph{Processes}, vol.~5, no.~3,
  p.~46, 2017.

\bibitem{lillicrap2015continuous}
T.~P. Lillicrap, J.~J. Hunt, A.~Pritzel, N.~Heess, T.~Erez, Y.~Tassa,
  D.~Silver, and D.~Wierstra, ``Continuous control with deep reinforcement
  learning,'' \emph{arXiv preprint arXiv:1509.02971}, 2015.

\bibitem{osti_6352074}
F.~McQuiston and J.~Parker, ``Heating, ventilating, and air conditioning:
  analysis and design.''

\bibitem{american1992ashrae}
A.~society of~heating refrigerating and air~conditioning engineers,
  \emph{{ASHRAE STANDARD}: An American Standard: Thermal Environmental
  Conditions for Human Occupancy}.\hskip 1em plus 0.5em minus 0.4em\relax
  American Society of Heating refrigerationg and air conditioning engineers,
  1992.

\bibitem{cheng2012thermal}
Y.~Cheng, J.~Niu, and N.~Gao, ``Thermal comfort models: {A} review and
  numerical investigation,'' \emph{Building and Environment}, vol.~47, pp.
  13--22, 2012.

\bibitem{van2012reinforcement}
M.~van Otterlo and M.~Wiering, ``Reinforcement learning and markov decision
  processes,'' in \emph{Reinforcement Learning}.\hskip 1em plus 0.5em minus
  0.4em\relax Springer, 2012, pp. 3--42.

\bibitem{mnih2015human}
V.~Mnih, K.~Kavukcuoglu, D.~Silver, A.~A. Rusu, J.~Veness, M.~G. Bellemare,
  A.~Graves, M.~Riedmiller, A.~K. Fidjeland, G.~Ostrovski \emph{et~al.},
  ``Human-level control through deep reinforcement learning,'' \emph{Nature},
  vol. 518, no. 7540, p. 529, 2015.

\bibitem{kummert2001optimal}
M.~Kummert, P.~Andr{\'e}, and J.~Nicolas, ``Optimal heating control in a
  passive solar commercial building,'' \emph{Solar Energy}, vol.~69, pp.
  103--116, 2001.

\bibitem{wang2000model}
S.~Wang and X.~Jin, ``Model-based optimal control of {VAV} air-conditioning
  system using genetic algorithm,'' \emph{Building and Environment}, vol.~35,
  no.~6, pp. 471--487, 2000.

\bibitem{ma2012model}
Y.~Ma, F.~Borrelli, B.~Hencey, B.~Coffey, S.~Bengea, and P.~Haves, ``Model
  predictive control for the operation of building cooling systems,''
  \emph{IEEE Transactions on control systems technology}, vol.~20, no.~3, pp.
  796--803, 2012.

\bibitem{wei2014co}
T.~Wei, Q.~Zhu, and M.~Maasoumy, ``Co-scheduling of {HVAC} control, {EV}
  charging and battery usage for building energy efficiency,'' in
  \emph{Proceedings of the 2014 IEEE/ACM International Conference on
  Computer-Aided Design}.\hskip 1em plus 0.5em minus 0.4em\relax IEEE Press,
  2014, pp. 191--196.

\bibitem{oldewurtel2010energy}
F.~Oldewurtel, A.~Parisio, C.~Jones, M.~Morari, D.~Gyalistras, M.~Gwerder,
  V.~Stauch, B.~Lehmann, and K.~Wirth, ``Energy efficient building climate
  control using stochastic model predictive control and weather predictions,''
  in \emph{Proceedings of the 2010 American control conference}, no.
  EPFL-CONF-169733.\hskip 1em plus 0.5em minus 0.4em\relax Ieee Service Center,
  445 Hoes Lane, Po Box 1331, Piscataway, Nj 08855-1331 Usa, 2010, pp.
  5100--5105.

\bibitem{anderson2004robust}
C.~W. Anderson, D.~Hittle, M.~Kretchmar, P.~Young, J.~Si, A.~Barto, W.~Powell,
  and D.~Wunsch, ``Robust reinforcement learning for heating, ventilation, and
  air conditioning control of buildings,'' \emph{Handbook of Learning and
  Approximate Dynamic Programming}, pp. 517--534, 2004.

\bibitem{pan2015internet}
J.~Pan, R.~Jain, S.~Paul, T.~Vu, A.~Saifullah, and M.~Sha, ``An internet of
  things framework for smart energy in buildings: designs, prototype, and
  experiments,'' \emph{IEEE Internet of Things Journal}, vol.~2, no.~6, pp.
  527--537, 2015.

\bibitem{minoli2017iot}
D.~Minoli, K.~Sohraby, and B.~Occhiogrosso, ``{IoT} considerations,
  requirements, and architectures for smart buildings—energy optimization and
  next-generation building management systems,'' \emph{IEEE Internet of Things
  Journal}, vol.~4, no.~1, pp. 269--283, 2017.

\bibitem{foresee1997gauss}
F.~D. Foresee and M.~T. Hagan, ``Gauss-newton approximation to bayesian
  learning,'' in \emph{Proceedings of the 1997 international joint conference
  on neural networks}, vol.~3.\hskip 1em plus 0.5em minus 0.4em\relax
  Piscataway: IEEE, 1997, pp. 1930--1935.

\bibitem{hagan1994training}
M.~T. Hagan and M.~B. Menhaj, ``Training feedforward networks with the
  marquardt algorithm,'' \emph{IEEE transactions on Neural Networks}, vol.~5,
  no.~6, pp. 989--993, 1994.

\bibitem{silver2014deterministic}
D.~Silver, G.~Lever, N.~Heess, T.~Degris, D.~Wierstra, and M.~Riedmiller,
  ``Deterministic policy gradient algorithms,'' in \emph{ICML}, 2014.

\bibitem{uhlenbeck1930theory}
G.~E. Uhlenbeck and L.~S. Ornstein, ``On the theory of the brownian motion,''
  \emph{Physical review}, vol.~36, no.~5, p. 823, 1930.

\bibitem{trnsys_url}
``{TRNSYS}: Transient system simulation tool,'' \url{http://www.trnsys.com},
  accessed, October 2018.

\bibitem{pytorch_url}
``Pytorch: An open source deep learning platform,'' \url{https://pytorch.org},
  accessed, October 2018.

\bibitem{mysql_url}
``{MySQL:} the world's most popular open source database,''
  \url{https://www.mysql.com}, accessed, October 2018.

\bibitem{de1998global}
R.~J. De~Dear, ``A global database of thermal comfort field experiments,''
  \emph{ASHRAE transactions}, vol. 104, p. 1141, 1998.

\end{thebibliography}


%\newpage


%\vfill


\end{document}